\documentclass[reprint, prd, superscriptaddress, longbibliography, tightenlines, nofootinbib, eqsecnum, amsfonts, amsmath, amssymb, floatfix, notitlepage,twocolumn]{revtex4-1}
\pdfoutput=1

\usepackage[utf8]{inputenc}
\usepackage{euscript}
\usepackage{epsfig}
\usepackage{graphics}
\usepackage{graphicx}
\usepackage{amsmath}
\usepackage{amssymb}
\usepackage{bm}
\usepackage[usenames,dvipsnames,svgnames,table]{xcolor}
\usepackage{xspace}
\usepackage{times}
\usepackage{appendix}
\usepackage{lipsum}
\usepackage[nolist,nohyperlinks]{acronym}
\usepackage{float}
\usepackage{simplewick}
\usepackage{tabularx}
\usepackage{booktabs}
\usepackage{natbib, ifthen}
\usepackage{hyperref}
\usepackage{comment}
\usepackage{dsfont}
\usepackage[caption=false]{subfig}
\usepackage{soul}

\newcommand{\nlth}{\ensuremath{N_L^{(3/2)}}\xspace}
\newcommand{\nlzero}{\ensuremath{N_L^{(0)}}\xspace}
\newcommand{\rdnlzero}{\ensuremath{\mathrm{RD{\text -}}N_L^{(0)}}\xspace}

\newcommand{\nlone}{\ensuremath{N_L^{(1)}}\xspace}

\newcommand{\lcdm}{\ensuremath{\mathrm{\Lambda CDM}}\xspace}

\newcommand{\WL}{\ensuremath{\mathcal{W}_L}\xspace}
\newcommand{\Xdat}[0]{\ensuremath{ {X^{\rm dat}}}\xspace}
\newcommand{\cpp}[0]{\ensuremath{C^{\phi\phi}_L}\xspace}

\newcommand{\cppfid}[0]{\ensuremath{C^{\phi\phi, \rm fid}}\xspace}
\newcommand{\cpptrue}[0]{\ensuremath{C^{\phi\phi}_L(\theta)}\xspace}

\newcommand{\cpphqe}[0]{\ensuremath{C^{\hat \phi^{\rm QE} \hat \phi^{\rm QE}}_L}\xspace}

\newcommand{\hcpp}[0]{\ensuremath{\hat C^{\phi\phi}_L}\xspace}

\newcommand{\resp}[0]{\ensuremath{\mathcal{R}}\xspace}

\newcommand{\pMAP}[0]{\ensuremath{\phi^{\rm MAP}}\xspace}

\newcommand{\hpMAP}[0]{\ensuremath{\hat \phi^{\rm MAP}}\xspace}

\newcommand{\Cov}[0]{\ensuremath{\textrm{Cov}}\xspace} 
\newcommand{\av}[1]{\left \langle #1 \right \rangle}

\newcommand{\LM}[0]{{LM}}

\definecolor{ZurichBlue}{rgb}{.255,.41,.884} 		
\definecolor{ZurichRed}{rgb}{0.9, 0.1, 0} 			
\definecolor{ZurichGreen}{rgb}{.196,.504,.396} 		
\definecolor{ZurichYellow}{rgb}{1,.648,0} 			
\definecolor{dodgerblue}{rgb}{0.12, 0.56, 1.0}
\definecolor{azure}{rgb}{0.0, 0.5, 1.0}
\definecolor{awesome}{rgb}{1.0, 0.13, 0.32}
\definecolor{alizarincrimson}{rgb}{0.82, 0.1, 0.26}
\definecolor{mediumpurple}{rgb}{0.58, 0.44, 0.86}
\definecolor{lasallegreen}{rgb}{0.03, 0.47, 0.19}
\definecolor{orange}{rgb}{1,0.5,0}

\newcommand{\Nzero}{N^{(0)}}
\newcommand{\None}{N^{(1)}}
\newcommand{\Ntwo}{N^{(2)}}

\graphicspath{{Figures_main_letter/}}

\begin{document}

\title{Lensing power spectrum of the Cosmic Microwave Background with deep polarization experiments}

\author{Louis Legrand }\email{louis.legrand@unige.ch}
\affiliation{Universit\'e de Gen\`eve, D\'epartement de Physique Th\'eorique et CAP, 24 Quai Ansermet, CH-1211 Gen\`eve 4, Switzerland}

\author{Julien Carron}\email{julien.carron@unige.ch}
\affiliation{Universit\'e de Gen\`eve, D\'epartement de Physique Th\'eorique et CAP, 24 Quai Ansermet, CH-1211 Gen\`eve 4, Switzerland}

\date{\today}

\begin{abstract}
    Precise reconstruction of the cosmic microwave background lensing potential can be achieved with deep polarization surveys by iteratively removing lensing-induced $B$ modes.
    We introduce a lensing spectrum estimator and its likelihood for such optimal iterative reconstruction. Our modelling share similarities to the state-of-the-art likelihoods for quadratic estimator-based (QE) lensing reconstruction.
    In particular, we generalize the $\Nzero$ and $\None$ lensing biases, and design a realization-dependent spectrum debiaser, making this estimator robust to uncertainties in the data modelling. We demonstrate unbiased recovery of the cosmology using map-based reconstructions, focussing on lensing-only cosmological constraints and neutrino mass measurement in combination with CMB spectra and acoustic oscillation data. We find this spectrum estimator is essentially optimal and with a diagonal covariance matrix.
    For a CMB-S4 survey, this likelihood can double the constraints on the lensing amplitude compared to the QE on a wide range of scales, while at the same time keeping numerical cost under control and being robust to errors.

\end{abstract}


\maketitle

\section{Introduction}

By probing the matter density fluctuations up to the last scattering surface, gravitational lensing of the Cosmic Microwave Background (CMB) is a powerful probe of the growth of structures, able to constrain the $\Lambda$CDM cosmological model, the neutrino mass scale, or modified gravity theories\cite{Lewis:2006fu}.
State-of-the-art reconstructions of CMB lensing maps \cite{Sherwin:2016tyf, Omori:2017tae, Planck:2018lbu, Wu:2019hek} have been obtained with a quadratic estimator (QE) \cite{Okamoto:2003zw}, which uses the anisotropic signatures of local shear and magnification distortions in the two-point function of the CMB. 
Optimal for current experimental noise levels, the QE will become inefficient for high-resolution, next generation survey data, such as CMB-S4 \cite{CMB-S4:2016ple}.  Since the primordial B-mode signal is small, it is well known that in principle one may reconstruct the lensing signal almost perfectly from the observed polarization \cite{Hirata:2003ka}, provided noise and foregrounds levels are put well below the lensing B-mode power of $\sim 5\mu \textrm{K}$-arcmin. In this regime, likelihood-based methods will be able to greatly improve the lensing reconstruction by using effectively higher order statistics of the CMB fields \cite{Hirata:2002jy, Hirata:2003ka, Millea:2017fyd, Millea:2020cpw, Carron:2017mqf, Carron:2018lcr}.

The QE power spectrum is a four-point function of the CMB maps. In addition to the sought-after lensing signal power \cpp, it contains other contributions, or `biases', that can be characterized analytically. The dominant bias, \nlzero, is due to Gaussian (disconnected) contractions of the four CMB data maps \cite{Hu:2001kj}, with a contribution from both the instrumental noise and CMB spectra. The next order bias (in an approach perturbative in the lensing power) is \nlone. It is due to the non-Gaussian secondary trispectrum contractions of the CMB fields created by lensing, and is proportional to \cpp~\cite{Kesden:2003cc}. For a perfectly Gaussian lensing map, the next order term is $\Ntwo_L$ which can be made negligible using suitable QE weights \cite{Hanson:2010rp}. The QE lensing map power spectrum can then be written schematically as
\begin{equation}\label{eq:Cpp}
   \cpphqe \sim \cpp + \nlzero + \nlone \; .
\end{equation}
This neglects the large-scale structure and post-born bispectrum, which source an additional bias,~$\nlth$. As shown in Refs.~\cite{Bohm:2016gzt, Bohm:2018omn, Beck:2018wud, Fabbian:2019tik}, this bias is generally expected to be quite small in the QE reconstruction from polarization which we consider here. Eq.~\eqref{eq:Cpp} then suggests that an estimate of the lensing power can be obtained by adequate subtraction of the bias terms. For parameter inference this is only slightly more complicated in practice, since the model-dependence of the QE response and of the biases on the CMB power spectra must be taken into account for the construction of an adequate spectrum likelihood \cite[see e.g.][]{Planck:2018lbu}. 

A lensing map reconstructed with a likelihood-based approach is a highly complicated function of the data, and it is analytically out of reach to track  the contributions to its power spectrum in a systematic manner. In this paper we investigate the lensing power spectrum from the optimal lensing map reconstruction developed in \cite{Carron:2017mqf}, showing that it shares a structure similar enough to the QE spectrum in Eq.~\eqref{eq:Cpp} that it is possible to use the very same type of likelihood construction to perform unbiased parameter inference. While the biases in this case are not four-point statistics of the data, they can nevertheless be accurately obtained with very similar calculations.

Compared to recent early attempts at Monte-Carlo sampling the lensing power~\cite{Millea:2020cpw, Millea:2020iuw}, this provides an absolutely massive reduction in numerical cost while reaching the expected improvements of likelihood-based methods as pioneered by Ref.~\cite{Hirata:2003ka}. We demonstrate this by performing parameter inference within $\Lambda$CDM and $\Lambda$CDM $+ \sum{m_\nu}$ models using optimal lensing map spectra as reconstructed on the full-sky with curved-sky geometry. Our framework is also directly applicable  to small and large sky areas inclusive of real-world non-idealities like sky cuts or inhomogeneous noise. The inner workings of this first curved-sky iterative reconstruction code making this possible will be presented elsewhere~\cite{CMBS4:inprep}.

State-of-the-art CMB lensing reconstructions make use of a robust, realization-dependent (RD) subtraction method of the leading noise bias, which uses QE's built partially on the data and partially on simulations \cite{Namikawa:2012pe, Story:2014hni, Ade:2015zua}. This has two main benefits. On one hand the subtraction of the bias is more accurate, as it removes mismatch between the data and fiducial model at first order in CMB spectra. This is often of great relevance, if only because the noise properties of CMB data are in some cases only crudely under control. 
On the other hand, this also reduces the covariance matrix of the spectrum estimate, which otherwise typically shows large positive correlations on small scales~\cite{Hanson:2010rp, Peloton:2016kbw}, as well as higher variance. An important ingredient that we introduce here is a generalization of this realization-dependent bias ($\rdnlzero$) which is suitable for the iterative estimate.

\section{Iterative Lensing spectrum estimator}
We start by reviewing briefly the maximum a posteriori (MAP) lensing reconstruction and discuss its normalization. We follow the algorithm of Ref.~\cite{Carron:2017mqf}, which has been demonstrated to work on data~\cite{POLARBEAR:2019snn}. We then present our analytical predictions of the responses and biases, including $\rdnlzero$.
\subsection{Estimator and normalization}

Let $\Xdat$ be an observed lensed CMB field (in what follows the Stokes parameters Q or U) including instrumental noise, and $\phi$ the lensing potential field (we neglect the small curl component here~\cite{Hirata:2002jy}). Assuming the unlensed CMB and noise fields are Gaussian,  we can define the log-likelihood
\begin{equation}
   \ln \mathcal{L}( \Xdat | \phi) = - \frac{1}{2}\Xdat^\dag \Cov_\phi^{-1}\Xdat - \frac{1}{2} \ln \det \Cov_\phi \; , 
\end{equation}
where $\Cov_\phi$ is the observed CMB (so including noise) covariance for fixed lenses, which can be modelled (to some workable approximation at least) using a fiducial model for the data noise, beam and transfer function, as well as fiducial unlensed CMB spectra with lensing and delensing operators at a given $\phi$ \cite[see][for more details]{Carron:2017mqf}. Using a Gaussian prior on the lensing potential with fiducial power spectrum $\cppfid_L$, we obtain the log-posterior
\begin{equation}
    \label{eq:logpost}
        \ln \mathcal{P}(\phi | \Xdat) = \ln \mathcal{L}( \Xdat | \phi) - \frac{1}{2}\sum_{\LM}  \frac{\phi_{\LM} \phi_{LM}^\dagger}{\cppfid_L}  \; .   
\end{equation}
The MAP lensing estimate $\hpMAP$ is defined as the one maximizing this posterior. Since the prior gradient is proportional to $\phi$ we can rearrange and write\footnote{Formally, $2\delta/\delta \phi^\dagger$ here means the harmonic transform of the variation $\delta /\delta \phi(\hat n)$ involving only real-valued variables.}
\begin{equation}\label{eq:master}
\begin{split}
	\hpMAP_{\LM} &= \cppfid_L  \left.\frac{2\delta \ln \mathcal L}  {\delta \phi_{\LM}^\dagger} \right|_{\phi = \hpMAP}
	\\
	&=\cppfid_L \left( g_{\LM}^{\rm QD} - \av{g_{\LM}^{\rm QD}}_{\hat \phi^{\rm MAP}} \right)
\end{split}
\end{equation}

where we introduced the quadratic gradient piece $g^{\rm QD}$ 
\begin{equation}
    \label{eq:gradients_2}
    \begin{split}
        g^{\rm QD}_\LM &= - \Xdat^\dag \Cov_{\hpMAP}^{-1}\frac{\delta \Cov_{\hpMAP} }{\delta \phi^\dagger_\LM} \Cov_{\hpMAP}^{-1}\Xdat \; .\\
    \end{split}
\end{equation} 
In \eqref{eq:master}, its average $\av{g_{\LM}^{\rm QD}}_{\hat \phi^{\rm MAP}}$, which is also the first variation of the log-determinant term, is the mean field, just like in the traditional QE analysis, with the difference that the deflection field is fixed to $\hpMAP$ in the average over realizations of the fiducial model. Explicit expressions for the quadratic piece in terms of spin-weight harmonic transforms are given in appendix A of~\cite{Carron:2017mqf}.

The quadratic gradient piece has implicit dependence on $\hpMAP$ through the delensing operators involved in the inverse variance filtering step $\Cov_{\hpMAP}^{-1}$, applied to the data maps. Nevertheless, Eq.~\ref{eq:master} shows that the MAP estimate truly \emph{is} quadratic in these delensed data maps.

Consider now the response to the true lenses 
\begin{equation}\label{eq:Wdef}
    \begin{split}
        \mathcal W_{LM L'M'} &\equiv \frac{\delta \hat \phi^{\rm MAP}_{LM}}{\delta \phi_{L'M'}} .
    \end{split}
\end{equation}
Since we are maximizing a posterior rather than a likelihood, we expect $\mathcal W$ to be a Wiener-filter: unity on resolved scales and suppressed elsewhere. 
Using Eq.~\eqref{eq:master}, we can write:
\begin{equation}
\begin{split}\label{eq:wf_split}
	\mathcal W_{LM L'M'}  &= \cppfid_L \Big[ \frac{\delta g^{\rm QD} }{ \delta \Xdat} \frac{\delta \Xdat }{\delta \phi} + \frac{\delta^2 \ln \mathcal L }{\delta \phi^\dagger \delta\phi} \frac{\delta \pMAP }{ \delta \phi} \Big]_{LM, L'M'} \;. 
\end{split}
\end{equation}
The left term of the above equation corresponds to the response of the estimator to the true lensing potential field. This is the direct analog to the standard QE response, for which we use similar notation $\mathcal R$.
The right term comes from the implicit dependency on $\pMAP$. This is equal to $-\mathcal H \mathcal W$, with $\mathcal H$ the log-likelihood Hessian curvature matrix. This results in the matrix equation $\mathcal W = \cppfid \left(\mathcal R - \mathcal H \mathcal W\right)$, or
\begin{equation}
\begin{split}\label{eq:signal}
	\mathcal W &= \left[ \frac{1}{\cppfid}  + \mathcal H  \right]^{-1} \mathcal R \; . \\
\end{split}
\end{equation}
In the regime where the prior is irrelevant, $1/ \cppfid$ becomes negligible in front of $\mathcal{H}$, and assuming that the reconstruction must be unbiased, we can write $\mathcal H^{-1} \mathcal R \sim \mathds{1}$ the identity matrix.
For practical purposes we may tentatively work with a simple isotropic approximation: 
\begin{equation}
\begin{split}
    \label{eq:wf_fid}
	\mathcal W_L & \equiv \left[\frac {1}{\cppfid_L} + \mathcal R_L   \right]^{-1}\mathcal R_L = \frac{\cppfid_L}{\cppfid_L + 1/ \mathcal R_L   } \; . \\
\end{split}
\end{equation}
This is indeed a Wiener-filter with fiducial  noise the inverse response $N_L = 1/\mathcal R_{L}$. This relation stands in direct analogy to the standard QE, where the reconstruction noise level $N^{(0)}$ is the exact inverse of the estimator response, provided the estimator weights are chosen to be optimal and the fiducial noise and CMB model match perfectly that of the data~\cite{Maniyar:2021msb}. We stick to the notation $\mathcal R$ instead of $N$ to emphasize that it behaves like a response, independent of the data noise maps and true lenses, as we empirically confirm in the next section.\\

\subsection{Likelihood construction, response and bias calculations}
\label{sec:resp_lik}

We use as our lensing data-vector the normalized raw spectrum 
\begin{equation}
    \hat C_L^{\phi\phi} = C_L^{\hpMAP \hpMAP} / \mathcal W^{2, \rm fid}_L \; , 
\end{equation}
where $C_L^{\hpMAP \hpMAP}$ is the pseudo power spectrum of the estimated $\hpMAP$, and $\mathcal W^{2, \rm fid}_L$ is our squared fiducial Wiener filter as defined by Eq.~\ref{eq:wf_fid}. 
To build the spectrum likelihood we must consider the case where the fiducial cosmology assumed to reconstruct the lensing map, denoted by the superscript $\rm fid$, is different from the sampled cosmology, denoted by $\theta$. The negative log-likelihood is
\begin{equation}
    \label{eq:likelihood}
    \begin{split}
        - 2 \ln L(\theta) =\left(\hat C_L^{\phi\phi} - C_L^{\, \rm pred}(\theta) \right) \Cov_{LL'}^{-1} \left(\hat C_{L'}^{\phi\phi} - C_{L'}^{\, \rm pred}(\theta) \right) \;,
    \end{split}
\end{equation}
where
\begin{equation}
    \label{eq:cpppred}
    \begin{split}
        C_L^{\rm pred}(\theta) &= \frac{\mathcal{W}_L^2(\theta)}{\mathcal{W}_L^{2, \rm fid}} \left[ C^{\phi \phi}_L(\theta) + N^{(1)}_L(\theta)  \right] \\
        &+\frac{\mathcal{W}_L^2(\theta)}{\mathcal{W}_L^{2, \rm fid}}\frac{\mathcal{R}_L^{2, \rm fid}}{\mathcal{R}_L^2(\theta)}\rdnlzero \; ,
    \end{split}
\end{equation}
and the covariance matrix $\Cov_{LL'}$ is estimated from simulations (see Sec.\ref{sec:param_est}).

The response $\resp_L$ and the \nlzero and \nlone biases are obtained by assuming that they follow the same analytical expressions of the standard QE, but replacing the lensed CMB and lensing convergence spectra in the weights of the estimator by partially delensed CMB and lensing convergence spectra. We proceed as follows.
Starting from fiducial and sampled lensing spectra, $\cppfid_L$ and $\cpptrue$, and fiducial and sampled unlensed CMB spectra $C_\ell^{EE, \rm fid}$ and $C_\ell^{EE}(\theta)$, we iteratively compute the partially-delensed lensing spectra and the corresponding partially-lensed CMB spectra. We iterate the following three steps:
\begin{enumerate}
    \item compute the partially-lensed (at the very first iteration fully-lensed) CMB spectra for the fiducial and sampled models, together with the non-perturbative `grad-lensed' spectra\footnote{The response function is computed with the grad-(partially)lensed spectra (defined as the cross-spectra of the CMB fields with their gradient, see Appendix C. of \cite{Lewis:2011fk}) which provide the most precise non-perturbative estimate of the response of the CMB spectra to lensing~\cite{Fabbian:2019tik}. In the case of reconstruction from polarization considered here usage of these is only a tiny correction to the (partially-)lensed spectra however.}, using the partially-lensed (at first fully lensed) lensing spectra. 
	\item calculate a QE reconstruction noise level $N_L^{(0)}$, using the fiducial partially-lensed spectra as QE weights and the sampled partially-lensed spectra in the lensing response. We turn this into a cross-correlation coefficient  $\rho_L = \cppfid_L / (\cppfid_L + \nlzero)$ of the lensing tracer.
	\item from $\rho_L$, update the fiducial and sampled partially-delensed lensing deflection spectra, given by $(1-\rho_L^2)\, \cppfid_L$ and $(1-\rho_L^2)\, \cpptrue$. We implicitly assume that this is what the MAP reconstruction is achieving.
\end{enumerate}
This procedure converges after a handful of steps. We can then calculate final estimates of the unnormalized $N^{(0)}$ and $N^{(1)}$ biases, for any choice of fiducial and sampled spectra. The response $\mathcal R_L(\theta)$ is calculated in the same way, using $C_\ell^{EE}(\theta)$ but always $\cppfid_L$ as lensing potential input. Our procedure is similar to the approaches of Ref.~\cite{Smith:2010gu, Hotinli:2021umk}.

Using the MAP estimate and the fiducial model assumed in the reconstruction procedure, we build simulations of the observed CMB where the unlensed CMB has been deflected by $\nabla \hat \phi^{\rm MAP}$, and the noise maps follow the fiducial noise model. This forms a set of simulation labeled `\emph{s}'. We then build quadratic estimates in the same way as the likelihood gradients $g^{\rm QD}$ (inclusive of the presence of $\hat \phi^{\rm MAP}$ in the filters), with the difference that on one of the two legs we use the actual data map, and on the second one a simulation. This gives a set of `estimates' $g^{ds_i}$, where by construction the response to lensing has been suppressed. We also form similar estimates $g^{s_i s_j}$ with a simulation on one leg, and another independent simulation (but still using the same deflection field $\nabla \hat \phi^{\rm MAP}$) on the second leg.  We then take the the auto pseudo spectra of these two sets, noted $\hat C_L^{di}$ and $\hat C_L^{ij}$ respectively. The combination
\begin{equation}
	\rdnlzero \equiv \frac{1}{\mathcal R^2_L}\av{4 \hat C_L^{di}- 2 \hat C_L^{ij} }_{\rm{MC's}}
\end{equation}
is then our realization-dependent $N^{(0)}$ noise estimate. This \rdnlzero is used to debias the power spectrum: we only need to compute it for the final lensing estimate \pMAP, and not at each step of the iterative lensing reconstruction.

We apply two empirical percent-level corrections, one to the Wiener-filter and one to the response $\mathcal R_L$, using fiducial simulations (see below Sec.~\ref{sec:results_spectra}), and we neglect any cosmology dependence of these corrections.
The parameter dependency of $\mathcal W$ in the first line of \ref{eq:cpppred} and the prefactor to \rdnlzero in the second line only enter through $C_\ell^{EE}(\theta)$, which is very tightly constrained by the CMB data, making their variation a tiny correction across a realistic MCMC chain. Furthermore, provided $\rdnlzero$ is close to the fiducial, these two terms cancel to a very large extent. Larger is the $C_L^{\phi\phi}(\theta)$ dependence that enters $N^{(1)}$ correction, but this is also a small effect in most relevant cases. By looking at reconstructions from simulations with intentionally grossly exaggerated deviations we can sanity check our implementation. To speed up computations while sampling the likelihood, we linearize around the fiducial model for the Wiener-filter, the response and the \nlone bias in $C_L^{\phi\phi}(\theta)- C_L^{\phi\phi, \rm fid}$ and $C_\ell^{EE}(\theta)-C_\ell^{EE, \rm fid}$, following the now standard approach of \cite{Ade:2015zua,Planck:2018lbu,  Simard:2017xtw, Sherwin:2016tyf} which we found perfectly fit for our purpose.

With this linearized likelihood, we can also produce lensing-only constraints by marginalizing out the uncertainty in the true CMB spectra~\cite{Planck:2018lbu}. This introduces small additional effects from using the observed realization of the CMB spectra to calculate the filter and responses, and of augmenting the covariance matrix by $ \sum_\ell \sigma^{2, E}_{\ell} \frac{\partial C^{\rm pred}_L}{ \partial C_\ell^{EE} } \frac{\partial C^{\rm pred}_{L'}}{ \partial C_\ell^{EE} } $, where $\sigma^{2, E}$ is the $E$-spectrum covariance, assumed here to be diagonal. In our configuration this results in a 2\% to 4\% increase on the lensing spectrum error bar at $L   \sim 700$ and $1400$ respectively, and less below that, so we do not consider it further in this paper.

\section{Results}
\subsection{Lensing power spectrum}
\label{sec:results_spectra}

\begin{figure}
    \centering
    \includegraphics[width= \columnwidth]{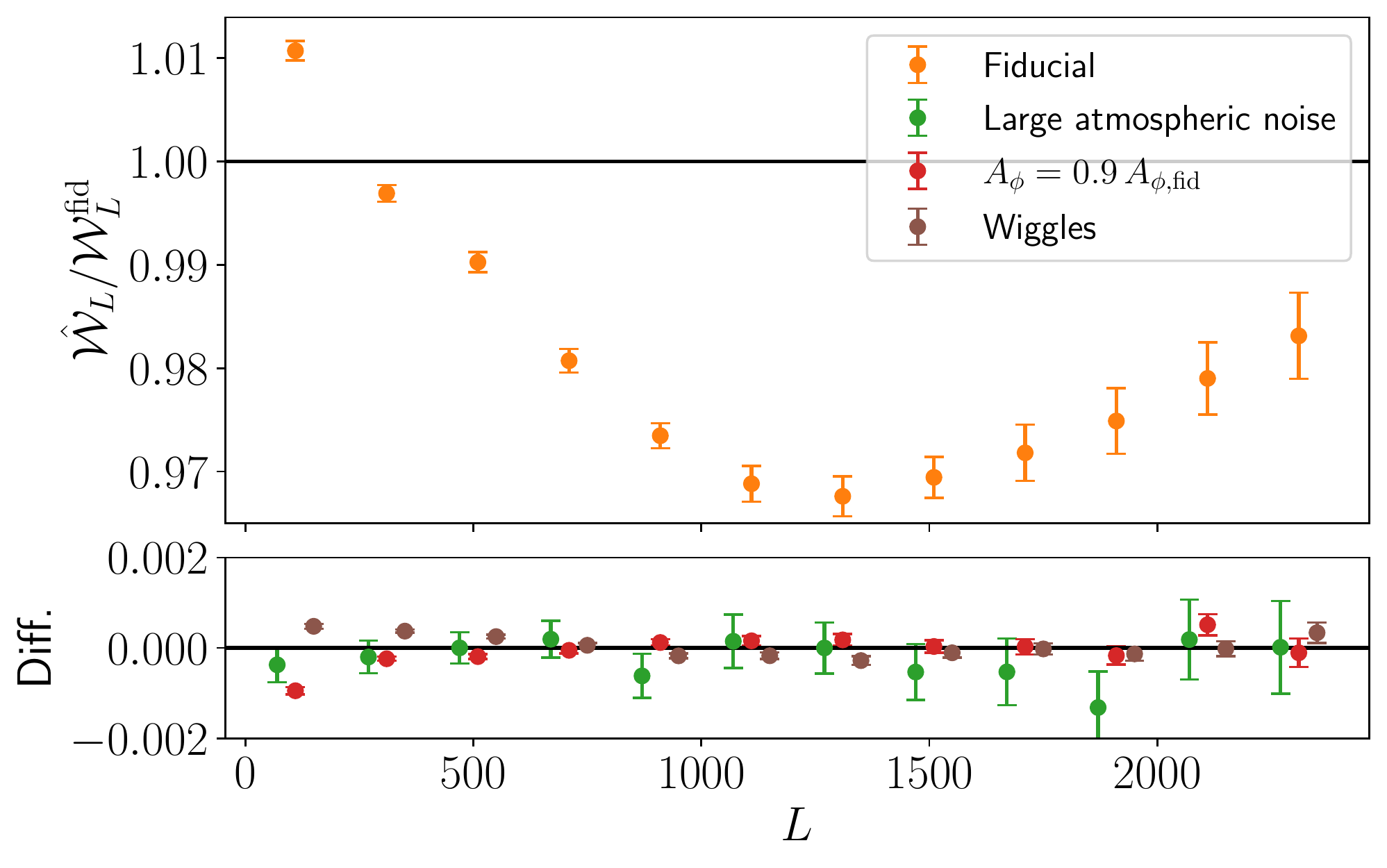}
    \caption{\textit{Upper panel:} Ratio of the empirical Wiener-filter, estimated from the cross correlation of the input lensing potential $\phi^{\rm in}$ and the estimated \hpMAP, over the fiducial Wiener-filter $\mathcal{W}_L^{\rm fid}$. Error bars show the standard error in each bin.
    Our analytical predictions provide a normalization of \hpMAP with a bias of at most $ \sim 3 \%$.  We use the bias estimated in this fiducial normalization to correct the \WL in the rest of our analysis, as justified by the lower panel. On the largest scales the empirical Wiener-filter is slightly larger than unity. This is due to our approximate treatment of the mean field which speeds up the computations but is otherwise inconsequential.
    \textit{Lower panel:} Difference between the \WL bias shown in the upper panel and three test cases discussed in the text: adding a large atmospheric noise component in the input map (green dots), having an input lensing spectrum lowered by 10\% (red dots) compared to the fiducial one, and when we multiply this input spectrum by a wiggling function (brown dots). In these three very extreme cases the \WL bias has less than 0.2\% difference from the fiducial case normalization bias. 
    }
    \label{fig:estimWF}
\end{figure}

We simulate several curved-sky, full sky CMB realizations with variations in the $\phi$ and $E$-mode inputs, with Gaussian noise corresponding roughly to CMB-S4 wide configuration with polarization noise level of $\Delta_P = \sqrt{2}~\rm \mu K arcmin$ and a beam FWHM of$~1~\rm arcmin$.
We start by checking empirically the Wiener-filter $\mathcal W$ of the $\hpMAP$ reconstructions, all performed using the same fiducial cosmology. We estimate an effective Wiener-filter from a cross-spectrum to the input lensing map
\begin{equation}
	\hat{\mathcal W}_L = \av{\frac{ C_L^{\phi^{\rm{in}} \hat \phi^{\rm MAP}}}{C_L^{\phi^{\rm{in}} \phi^{\rm in}}} }\; .
\end{equation}
Fig.~\ref{fig:estimWF} compares the empirical Wiener-filter to our prediction. The lower panel of this figure confirms two important and not a-priori obvious points making the spectrum likelihood tractable: the Wiener-filter is independent of both the actual data noise and of the true lensing signal. To show this clearly we have used some extreme deviations from the fiducial model. To illustrate the first point we have included a large atmospheric noise component in the simulated data map, corresponding to the green points, given by $N_\ell = \Delta^2_P \left(1 + \left(\frac{\ell}{700}\right)^{-1.4} \right)$, which is completely ignored in the covariance model of the reconstruction, using white noise power $\Delta_P^2$. To illustrate the second point, we have used simulations with tweaked input lensing potential power. The first has an amplitude decreased by 10\% (red points), in the second a strong oscillatory signal has been superimposed to the fiducial spectrum, of the form of a factor $(1 + 0.1 \sin(2 \pi \ell / 200))$ (brown points).

As already mentioned in the previous section, we correct empirically two analytical ingredients of the MAP spectrum using four independent simulations where the cosmology matches exactly the fiducial model of the reconstruction. The first correction is to the Wiener-filter, as shown in Fig.~\ref{fig:estimWF}. The second is to the response $\mathcal R$: we rescale $\mathcal R_L $ such that $\av{\rdnlzero} = 1 /{\mathcal R_L}$, as should hold for a QE with optimal weights.

\begin{figure} 
    \centering
    \includegraphics[width = \columnwidth]{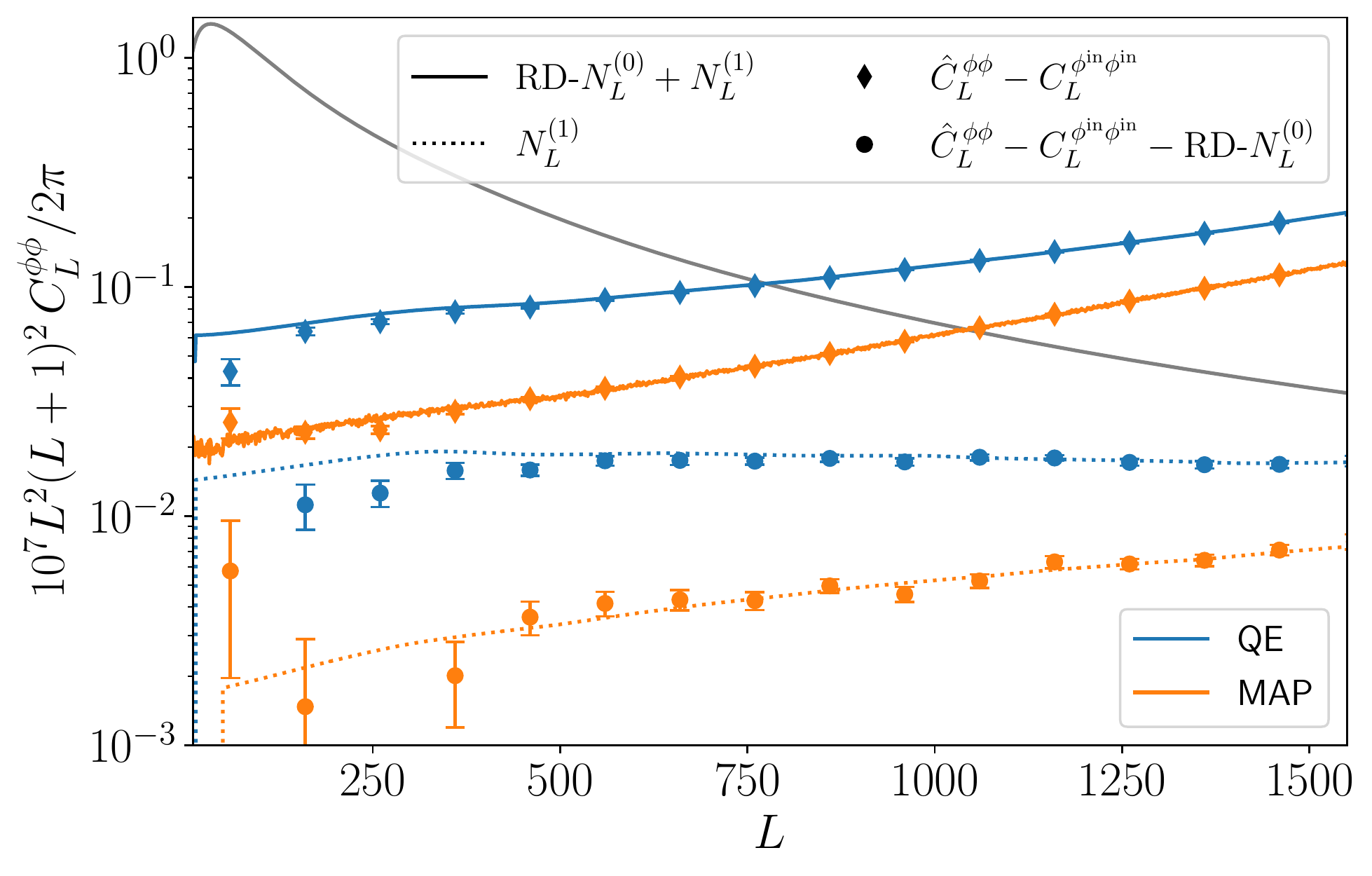}
    \caption{Estimated lensing spectra subtracted by the input potential map spectrum (suppressing cosmic variance) for the QE (blue diamonds) and MAP estimators (orange diamonds). These should correspond to the sum of the \rdnlzero and \nlone biases, shown as solid lines. Also shown are the estimated spectra subtracted by the input spectrum and by \rdnlzero (in blue and orange circles for the QE and the MAP respectively). These should be dominated by the \nlone bias, with our analytical predictions shown as the dotted lines. The grey line shows the fiducial lensing spectrum.
    }
    \label{fig:cpp}
\end{figure}

Fig.~\ref{fig:cpp} shows the estimated lensing spectra, for both QE and MAP estimators, after subtracting it by the spectrum of the input map (cancelling cosmic variance), as well as by the estimated \rdnlzero bias. This shows that the estimated \rdnlzero and \nlone are accurate predictions of the biases in the estimated lensing spectra. The $N^{(1)}$ bias predictions starts to be less accurate at the lowest lensing multipoles, possibly in part because they are calculated in the flat-sky approximation. Since $N^{(1)}$  is very small on these scales this is of no significant relevance. In our likelihoods the $N^{(1)}$ bias for the MAP is set to zero for $L<50$.
We also see that for $L\sim 800$, the \rdnlzero bias is reduced by a factor of two with the MAP, owing to the reduction in $B$-power achieved by the iterations. $\nlone$, also proportional to the residual lensing power, is suppressed even further.

\begin{figure}
    \centering
    \includegraphics[width = \columnwidth]{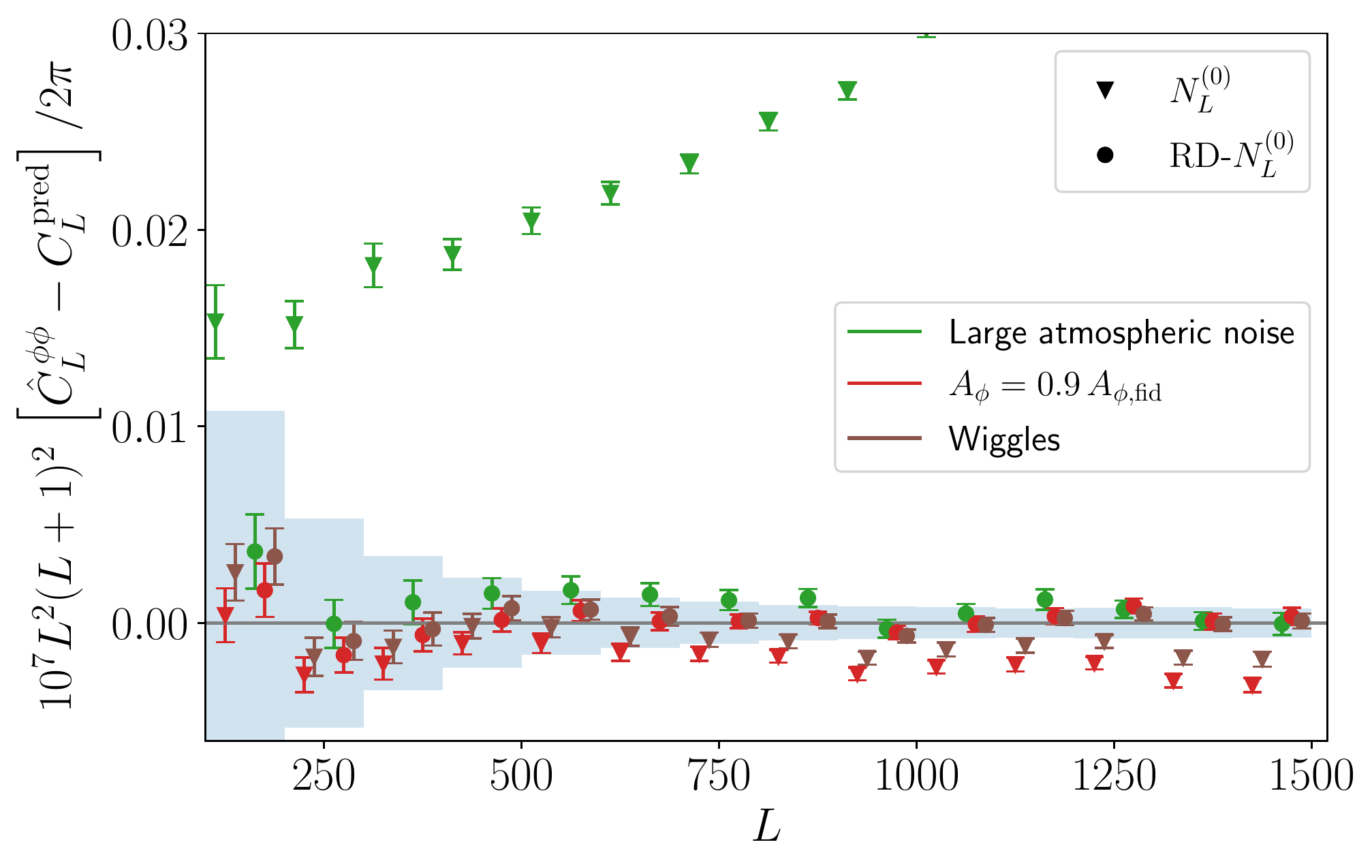}
    \caption{Biases between the estimated and predicted spectra debiased with the fiducial \nlzero(triangles) or using \rdnlzero(dots), for the same three non-fiducial, extreme realizations as on Fig.~\ref{fig:estimWF}. 
    The predicted spectra include the input spectrum of the map instead of the true spectrum, which cancels cosmic variance. We correct the response and \nlone bias at first order to the true lensing and polarization spectra, just as we do when sampling the cosmological model. The blue bars show our binning and one-realization statistical errors. 
    The $\sim 1 \sigma $ positive bias of the atmospheric noise case from $\sim 500  $ to $\sim 800$ is due to an increased \nlone with respect to the predictions, which is well captured by including the true noise in the computation of the \nlone bias instead of the fiducial one.
    } 
    \label{fig:rdn0_debias}
\end{figure}

The robustness of our predicted lensing spectrum $C_L^{\rm pred}$ with respect to the fiducial and true spectra is shown in Fig.~\ref{fig:rdn0_debias}. We estimate \hcpp for the same three extreme cases of Fig.~\ref{fig:estimWF}. These three spectra are all normalized with the same \WL corrected for the bias of the fiducial case. The predictions $C_L^{\rm pred}$ include either the \nlzero or the \rdnlzero and the \nlone biases. For illustrative purposes, we use here the lensing spectrum of the input maps to cancel cosmic variance instead of the true lensing spectrum.
In non-fiducial cases, the \nlzero debiaser cannot recover an unbiased prediction of the reconstructed spectrum. 
Debiasing with \rdnlzero and including the corrections at first order in $C_L^{\phi\phi}(\theta)- C_L^{\phi\phi, \rm fid}$ and $C_\ell^{EE}(\theta)-C_\ell^{EE, \rm fid}$ for the response and \nlone bias allows to get an accurate prediction of the estimated spectrum \hcpp.

\begin{figure}
    \includegraphics[width=\columnwidth]{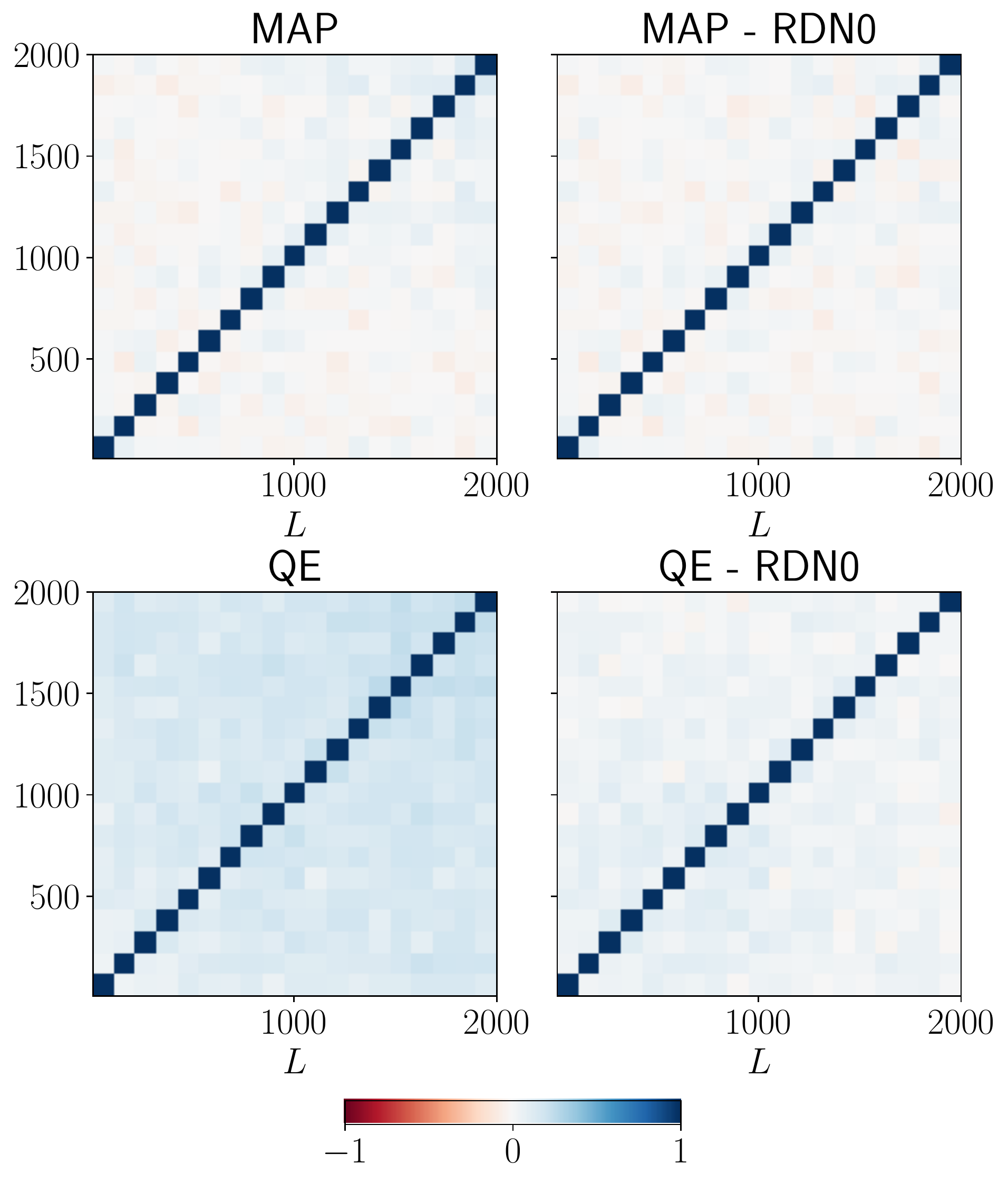}
    \caption{Correlation matrices of MAP (top row) and QE (bottom row) lensing convergence spectra from 1024 flat-sky simulations, binned with a step $\Delta L = 100$ between $L_{\rm min} = 10$ and $L_{\rm max} = 2000$. In the right panels the spectra are debiased with \rdnlzero while in the left panels they are not.}
    \label{fig:covmat}
\end{figure}

\begin{figure}
    \centering
    \includegraphics[width=\columnwidth]{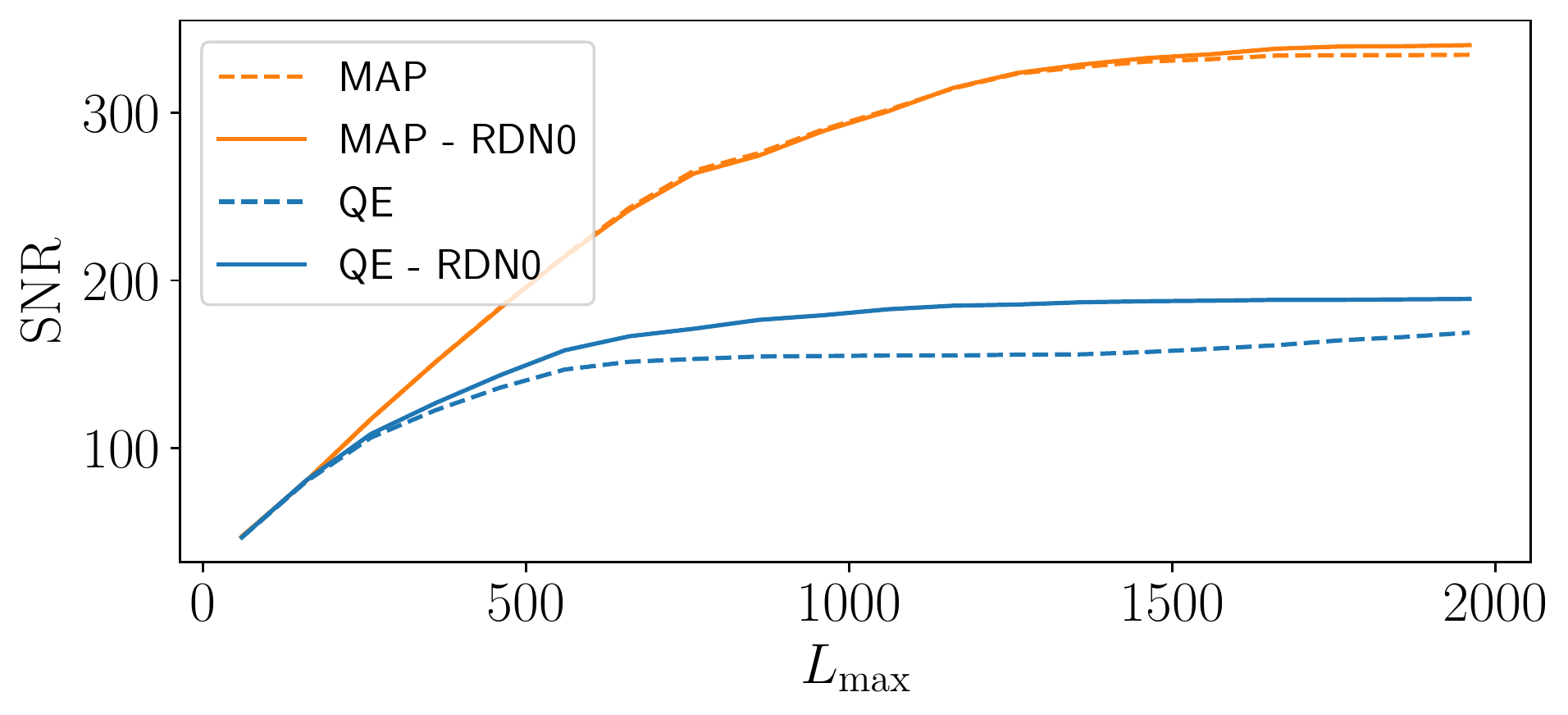}
    \caption{Signal to noise ratio for the MAP (orange lines) and the QE (blue lines) lensing power spectra. For dashed lines the spectra are debiased with \rdnlzero, while plain lines are not. The MAP increases the SNR by $80 \%$ compared to the QE, from 190 to 340.}   
    \label{fig:snr}
\end{figure}

To obtain the covariance matrix, we estimate the QE and MAP lensing spectra from 1024 flat-sky simulations, as well as their \rdnlzero. Fig.~\ref{fig:covmat} shows the correlation matrices (the covariance matrix normalized by the diagonal) for the QE and MAP spectra, with and without realization dependent debiasing. We see that, even without realization dependent debiasing, the MAP spectrum is less correlated between different multipole bins. The decrease of non-diagonal correlations using a realization debiaser seems to be negligible, and is only visible in the highest multipole bins ($L>1500$). It appears this reduction of non-diagonal correlations is less important in the MAP than it is for the QE spectrum.
On Fig.~\ref{fig:snr} we show the cumulative signal to noise ratio (SNR) given by
\begin{equation}
    \mathrm{SNR}(L_{\rm max}) =  \sqrt{\sum_{L_{\rm min}}^{L_{\rm max}} C_L^{\phi \phi, \rm fid}  \Cov_{LL'}^{-1} C_{L'}^{\phi \phi, \rm fid}} \; .
\end{equation}
The cumulative SNR of the QE with a realization dependent debiaser saturates around 190 at $L_{\rm max} = 1000$, while for the MAP the SNR keeps increasing up to $\sim 340$ at $L_{\rm max} = 1500$. The information gained with the use of the \rdnlzero is less important for the MAP than for the QE, because the covariance matrix of the MAP is already optimal in the range of scales which contain most of the signal.

\subsection{Cosmological parameter estimation}
\label{sec:param_est}

We demonstrate the potential of our MAP spectrum estimator to obtain constraints on cosmological models for a CMB-S4 experiment. 
First, using only our lensing likelihood we measure the constraints on the $\sigma_8 \Omega_{\rm m}^{0.27}$ parameter combination. Second, we let free the sum of neutrino masses and obtain constraints on it by adding CMB and baryonic acoustic oscillations (BAO) likelihoods. 
To check that our pipeline is unbiased, we simulate two datasets, each in a different cosmology. 
The first one is the \emph{Planck}~FFP10 model\footnote{\url{https://github.com/carronj/plancklens/blob/master/plancklens/data/cls/FFP10_wdipole_params.ini}}, with $\sum m_\nu = 0.06 \, \rm eV$. This is our fiducial cosmology for all lensing reconstructions.
The second one has an higher neutrino total mass of $\sum m_\nu = 0.1 \, \rm eV$, with three massive neutrinos in a normal hierarchy and two degenerate mass eigenstates, while keeping the angular size of the sound horizon fixed. The latter is performed by changing the Hubble constant slightly.
For both cosmologies we simulate five full-sky lensed CMB, with CMB-S4 noise level, and we estimate the MAP lensing spectrum from the polarization maps. 
These spectra are binned between $L_{\rm min} = 10$ and $L_{\rm max} = 1500$ with a step $\Delta L = 10$ up to $L=100$, and a step $\Delta L = 50$ above. 
The covariance matrix for $L>100$ is obtained from lensing spectra estimated with 1024 flat-sky simulations of $645~\rm deg^2$ each. This matrix is normalized in order to reproduce a 40~\% sky fraction. The larger scales, $L<100$, use an analytical Gaussian covariance with same sky fraction.
We sample our likelihoods with the code Cobaya \cite{Torrado:2020dgo}, using an adaptive MCMC sampler \cite{Lewis:2002ah,Lewis:2013hha}, and the Boltzmann code CAMB \cite{Lewis:1999bs,Howlett:2012mh}.

\begin{figure}
    \centering
    \includegraphics[width=\columnwidth]{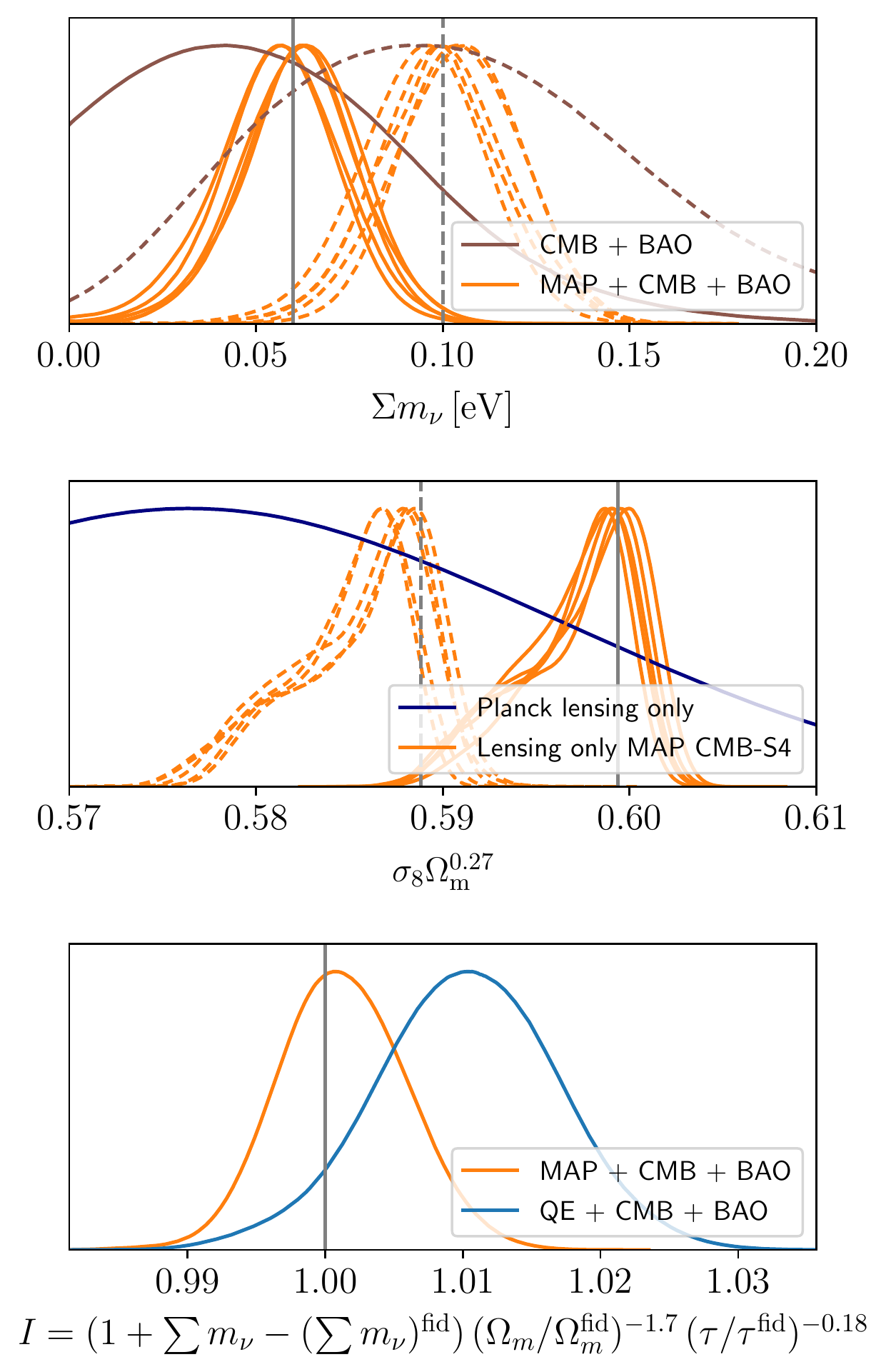}
    \caption{
        \textit{Upper panel:} $\Lambda$CDM $+\sum_\nu m_\nu$ marginalized posterior on the sum of neutrino masses for our test CMB-S4 lensing likelihoods. Constraints from combined CMB-S4 and DESI BAO are shown in brown, while constraints including further the MAP lensing spectrum are shown in orange. For solid lines the true neutrino mass is of 0.06 eV (solid grey line), corresponding to the fiducial cosmology in our lensing reconstruction. For dashed lines, the true neutrino mass is of 0.1 eV (dashed grey line) while the fiducial cosmology for the reconstruction stays the same. For each dataset we show the posterior of five different lensed CMB realisations. 
        \textit{Medium panel:} The orange lines show $\Lambda $CDM marginalized constraints on the CMB-lensing parameter $\sigma_8\Omega_{\rm m}^{0.27}$ with our MAP lensing-only likelihood for a CMB-S4 survey. The dark blue line shows the posterior from the latest \emph{Planck} data for comparison~\cite{Planck:2018lbu}. The solid, and dashed lines stands for the two inputs cosmologies, as above. The shape of the posterior is driven by the remaining correlations in the $\sigma_8-\Omega_{\rm m}$ plane.
        \textit{Lower panel:} $\Lambda$CDM $+\sum_\nu m_\nu$ marginalized posterior on the derived parameter $I$, for the combination of our lensing, CMB-S4 and DESI BAO likelihoods, in the fiducial dataset. The blue line uses a QE to estimate the lensing spectrum, while the orange line uses the MAP estimator. On this combined parameter the MAP spectrum is able to reduce the marginalised $1\sigma$ constraint by $30\%$.
    }
    \label{fig:mnu_mcmc}
\end{figure}

The sensitivity  of CMB lensing to \lcdm cosmological parameters is discussed in \cite{Pan:2014xua, Planck:2015mym, SPT:2019fqo}. There is a three parameter degeneracy $\sigma_8$ - $\Omega_{\rm m}$ - $H_0$ `tube', which projects onto a tightly constrained $\sim \sigma_8 \Omega_{\rm m}^{0.25}$.
For our lensing-only constraints, we use the same priors as the \emph{Planck} analysis~\cite{Planck:2018lbu}, most notably a prior on the baryon density from abundance measurements that constrains the sound horizon (a prior much weaker than the constraints expected from CMB-S4).
The marginalized posterior on the CMB lensing parameter are shown in the lower panel of Fig.~\ref{fig:mnu_mcmc}, where we also show for comparison the constraints from the \emph{Planck} lensing-only analysis \cite{Planck:2018lbu}. For both input cosmologies we recover an unbiased estimate of the $\sigma_8 \Omega_{\rm m}^{0.27}$ parameter combination, with constraints about seven times better than current best data (the 0.27 exponent was found with a principal component analysis of our chains).

CMB lensing is sensitive to the sum of neutrino masses through the suppression of the growth on scales smaller than the free streaming scale \cite{Lesgourgues:2006nd}. Combining primordial CMB spectra and BAO can break the CMB degeneracies by putting constraints on the sound horizon at low and high redshift~\cite{2dFGRSTeam:2002tzq}. Good knowledge of the optical depth to reionization $\tau$ fixes then the primordial fluctuation amplitude. Combining CMB + BAO + CMB lensing then provides a constraint on the neutrinos total mass. Current tightest constraints on the neutrino masses obtained by combining CMB + BAO + CMB lensing datasets are of $\sum m_\nu < 0.12 \, \rm eV$ (95\% confidence level) \cite{Planck:2018vyg}, and the CMB-S4 + DESI BAO + CMB-S4 lensing combination is expected to be able to detect to high significance the minimal neutrino mass allowed by terrestrial experiments~\cite{CMB-S4:2016ple}. 
We sample the seven parameters of the \lcdm$+\sum m_\nu$ model for the combined posterior including CMB and BAO likelihoods, with or without our MAP lensing likelihood. We include a Gaussian prior $\sigma_\tau =  0.002$. This corresponds roughly to the cosmic variance limit for a full sky polarization survey such as LiteBIRD \cite{LiteBIRD:2020khw}. Our simulated CMB data vector is a set of simulated TT, TE and EE CMB unlensed spectra  ranging from $\ell_{\rm min} = 30$ to $\ell_{\rm max} = 3000$, with covariance rescaled to 40\% of the sky observations. The BAO likelihood reproduces the one used to forecast the DESI survey \cite{DESI:2016fyo}, following the recipe of \cite{Font-Ribera:2013rwa}. We include the main galaxy sample, the low redshift bright galaxy sample, and the high redshift Lyman-$\alpha$ quasar survey, for a total redshift range from $z=0.05$ to $z=3.55$. The marginalized constraints on the sum of neutrino masses are shown in the upper panel of Fig.~\ref{fig:mnu_mcmc}. In both the low- and high-mass  cosmologies, our likelihood pipeline obtains an unbiased estimate of the true neutrino masses. We obtain a $3.8 \sigma$ (respectively $6.3 \sigma$ in the high mass case) detection of massive neutrinos, consistent with the forecasts presented in the CMB-S4 Science book \cite{CMB-S4:2016ple}.

While the total signal to noise ratio of the lensing spectrum with the MAP is 80\% higher than with the QE, the marginalized constraint on the sum of neutrino mass is only slightly improved.
We obtain $\sigma_{M_\nu}=0.016 \, \rm eV$ and  $\sigma_{M_\nu}=0.017 \, \rm eV$  for the MAP and QE respectively. This rather small difference is consistent with the Fisher analysis of \cite{Allison:2015qca}. 
This probably means that the marginalised constraint is not driven by the improvement in statistical power brought by the MAP estimator, but by degeneracies between cosmological parameters. To test this, we perform a principal component analysis of our chains on the parameters $\sum m_\nu, \Omega_{\rm m} $ and $\tau$, among the main parameters impacting the amplitude of the lensing power spectrum.
We found that the combined parameter $I= \left(1 + \sum m_\nu - (\sum m_\nu)^{\rm fid}\right) \left(\Omega_{\rm m} /\Omega_{\rm m}^{\rm fid}\right)^{-1.7} \left(\tau/\tau^{\rm fid}\right)^{-0.18}$, gets better constraints with the MAP than with the QE, as shown on the lower panel of  Figure \ref{fig:mnu_mcmc}. The $1\sigma$ constraints are of $\sigma_I=6.7 \, \rm meV$ and $\sigma_I=4.7 \, \rm meV$ with the QE and MAP respectively, corresponding to an improvement of $30 \%$. This does not yet match the statistical improvement obtained by the MAP reconstruction, so there might still be some degeneracies left.

\begin{figure}
    \centering
    \includegraphics[width=\columnwidth]{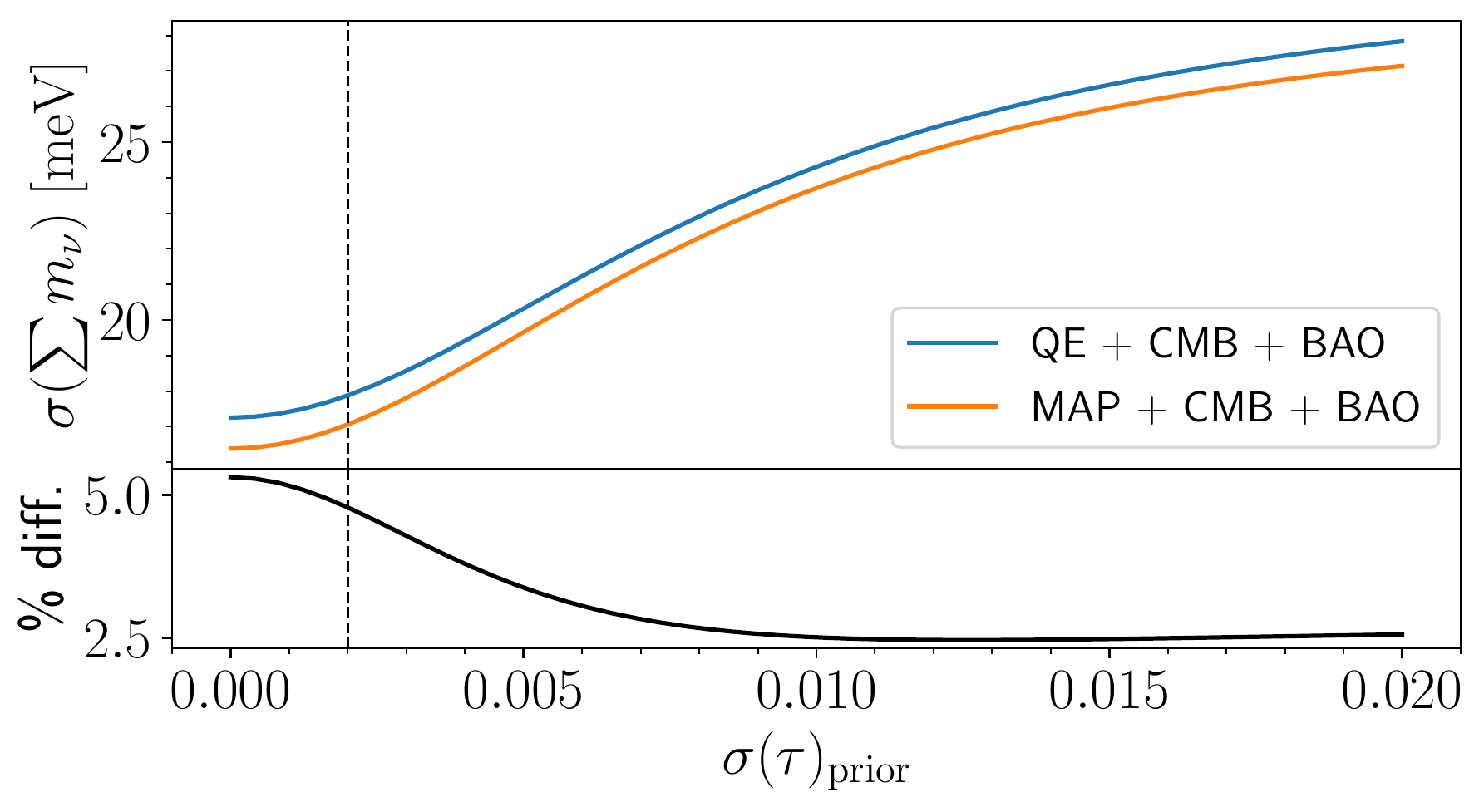}
    \caption{\textit{Upper panel:} Marginalized $1\sigma$ Fisher constraints on the sum of neutrino masses as a function of standard deviation of the Gaussian prior on $\tau$. Blue (orange) line combines the CMB-S4 and DESI BAO likelihoods with the QE (resp. MAP) lensing likelihood. The dashed line shows the prior of $\sigma(\tau) = 0.002$ we used in our MCMC analysis. \textit{Lower panel:} Improvement (in percent) on the neutrino mass constraint when using the MAP instead of the QE lensing spectrum.}
    \label{fig:tau}
\end{figure}

In order to assess the impact of the reionisation optical depth prior on the neutrino mass constraints, we compute the Fisher matrix for the combination of our CMB-S4 lensing potential, CMB-S4 unlensed and DESI BAO likelihoods as above. We use either the QE or the MAP lensing likelihoods. Fig.~\ref{fig:tau} shows the marginalized $1\sigma$ constraints on $\sum m_\nu$ as a function of the standard deviation of the Gaussian prior on $\tau$. When using the MAP lensing likelihood, we see that the constraint is of $\sigma_{M_\nu}=0.016 \, \rm eV$ when $\tau$ is fixed (i.e. $\sigma(\tau)_{\rm prior}=0$) and increases to $\sigma_{M_\nu}=0.027 \, \rm eV$ with a looser prior on $\tau$. This is in agreement with our constraints from the MCMC chains above (using the prior of $\sigma(\tau)_{\rm prior}=0.002$), and is also in agreement with the CMB-S4 forecasts \cite{CMB-S4:2016ple}. 
Here also we see that the improvement from the QE to the MAP constraints is of only a few percents, and at best of 5\% when $\tau$ is fixed. 
As we showed above, there are still degeneracies between cosmological parameters which have to be broken to reach the full statistical power of the MAP lensing reconstruction.

\section{Discussion}

In this paper we introduced a new CMB lensing power spectrum estimator, using optimal likelihood-based lensing reconstruction which carries much more signal to noise for experiments with low polarization noise. Keeping the numerical cost under control, in principle directly applicable to masked data or with other non-idealities, with a robust realization-dependent debiaser, this essentially solves several practical challenges facing reconstruction of the lensing power spectrum beyond the QE.

Our spectrum estimator uses altogether a single MAP lensing reconstruction, performed assuming a fiducial cosmological model. While the MAP reconstruction is much more expensive than its QE counterpart and dominates the overall numerical cost, the entire pipeline remains all things considered quite economical. As a point of comparison to the literature, the recent iterative spectrum estimation proposal of Ref.~\cite{Millea:2021had} performs of $O(10^3)$ MAP reconstructions. Owing to the somewhat arbitrary choice of fiducial cosmology, our estimator will be somewhat sub-optimal. However, our cosmological model is so tightly constrained already at the present time, that this can hardly be more than a percent-level effect.

We neglected several issues. CMB foregrounds (such as Galactic dust emission, Sunyaev-Zel’dovich effect, radio sources or the Cosmic Infrared Background), particularly in temperature, can bias the lensing reconstruction by creating non-Gaussian signatures in the observed CMB \cite{vanEngelen:2013rla, Osborne:2013nna, Ferraro:2017fac, Madhavacheril:2018bxi, Schaan:2018tup, Mishra:2019qyd, Sailer:2020lal, Darwish:2021ycf,Lembo:2021kxc}. It is not known yet what is the size of these biases in the MAP reconstruction. We also neglected the bias from the large-scale bispectrum~\cite{Bohm:2016gzt, Bohm:2018omn,Beck:2018wud, Fabbian:2019tik}, more relevant on the smaller scales probed by CMB-S4 lensing than it is today. This bias is especially important when performing a tomographic analysis of CMB lensing in cross-correlation with galaxy surveys \cite{Hirata:2008cb, Schaan:2016ois, Schmittfull:2017ffw, DES:2018fvb,  Euclid:2021qvm}, a promising probe of the growth of structures. Here also it is not yet known what is the importance of this bias when using the MAP reconstruction. For these cross-correlations, our results will be useful to model accurately the normalization of the MAP estimate. Finally, we did not discuss the origin of the small corrections to the fiducial Wiener-filter and response. As we showed, this does not limit nor bias our analysis, but there is room for a more precise analytical understanding of our results. 

We did not consider masking and other instrumental non-idealities either in this work. The MAP reconstruction is known to work on realistic data. It is worth noting that all ingredients introduced here, inclusive of \rdnlzero, can be obtained in just the same way on a masked sky.  It is well known that close the the mask boundaries, the QE isotropic normalization is inaccurate, and must be corrected for by small Monte-Carlo correction. In a preliminary analysis, we found as expected the same behavior for the MAP reconstruction, with a bias less than a few percents. This bias is of similar size to the one for the full-sky analysis presented in this paper, and hence it is simple to account for them jointly. Another important difference in the masked case is the much larger size of the low-$L$ mean-field, induced by the large-scale anisotropies from notably the mask and scanning pattern.  An exact treatment of this mean-field would in principle require the analysis of a number of simulations at each iteration step, multiplying the numerical cost by this number. However, Ref.~\cite{POLARBEAR:2019snn} already demonstrated that one can make profit of the small size of the $\hat \phi$ dependence of the mean-field to use the QE mean-field estimate in each of the MAP iterations. For this reason we do not expect the mean-field to slow down the pipeline. We leave for future work the full adaptation of this analysis pipeline to more realistic simulations of the CMB-S4 survey (or other planned observations), or actual data.

\acknowledgements{The authors wish to thank Giulio Fabbian, Antony Lewis, Blake Sherwin and Kimmy Wu for useful discussions and comments, as well as the anonymous referees to the first version of this paper. The authors acknowledge support from a SNSF Eccellenza Professorial Fellowship (No. 186879).}

\bibliography{biblio_clean.bib}

\begin{thebibliography}{62}%
\makeatletter
\providecommand \@ifxundefined [1]{%
 \@ifx{#1\undefined}
}%
\providecommand \@ifnum [1]{%
 \ifnum #1\expandafter \@firstoftwo
 \else \expandafter \@secondoftwo
 \fi
}%
\providecommand \@ifx [1]{%
 \ifx #1\expandafter \@firstoftwo
 \else \expandafter \@secondoftwo
 \fi
}%
\providecommand \natexlab [1]{#1}%
\providecommand \enquote  [1]{``#1''}%
\providecommand \bibnamefont  [1]{#1}%
\providecommand \bibfnamefont [1]{#1}%
\providecommand \citenamefont [1]{#1}%
\providecommand \href@noop [0]{\@secondoftwo}%
\providecommand \href [0]{\begingroup \@sanitize@url \@href}%
\providecommand \@href[1]{\@@startlink{#1}\@@href}%
\providecommand \@@href[1]{\endgroup#1\@@endlink}%
\providecommand \@sanitize@url [0]{\catcode `\\12\catcode `\$12\catcode
  `\&12\catcode `\#12\catcode `\^12\catcode `\_12\catcode `\%12\relax}%
\providecommand \@@startlink[1]{}%
\providecommand \@@endlink[0]{}%
\providecommand \url  [0]{\begingroup\@sanitize@url \@url }%
\providecommand \@url [1]{\endgroup\@href {#1}{\urlprefix }}%
\providecommand \urlprefix  [0]{URL }%
\providecommand \Eprint [0]{\href }%
\providecommand \doibase [0]{http://dx.doi.org/}%
\providecommand \selectlanguage [0]{\@gobble}%
\providecommand \bibinfo  [0]{\@secondoftwo}%
\providecommand \bibfield  [0]{\@secondoftwo}%
\providecommand \translation [1]{[#1]}%
\providecommand \BibitemOpen [0]{}%
\providecommand \bibitemStop [0]{}%
\providecommand \bibitemNoStop [0]{.\EOS\space}%
\providecommand \EOS [0]{\spacefactor3000\relax}%
\providecommand \BibitemShut  [1]{\csname bibitem#1\endcsname}%
\let\auto@bib@innerbib\@empty
\bibitem [{\citenamefont {Lewis}\ and\ \citenamefont
  {Challinor}(2006)}]{Lewis:2006fu}%
  \BibitemOpen
  \bibfield  {author} {\bibinfo {author} {\bibfnamefont {Antony}\ \bibnamefont
  {Lewis}}\ and\ \bibinfo {author} {\bibfnamefont {Anthony}\ \bibnamefont
  {Challinor}},\ }\bibfield  {title} {\enquote {\bibinfo {title} {{Weak
  gravitational lensing of the cmb}},}\ }\href {\doibase
  10.1016/j.physrep.2006.03.002} {\bibfield  {journal} {\bibinfo  {journal}
  {Phys. Rept.}\ }\textbf {\bibinfo {volume} {429}},\ \bibinfo {pages} {1--65}
  (\bibinfo {year} {2006})},\ \Eprint {http://arxiv.org/abs/astro-ph/0601594}
  {arXiv:astro-ph/0601594 [astro-ph]} \BibitemShut {NoStop}%
\bibitem [{\citenamefont {Sherwin}\ \emph {et~al.}(2017)\citenamefont {Sherwin}
  \emph {et~al.}}]{Sherwin:2016tyf}%
  \BibitemOpen
  \bibfield  {author} {\bibinfo {author} {\bibfnamefont {Blake~D.}\
  \bibnamefont {Sherwin}} \emph {et~al.},\ }\bibfield  {title} {\enquote
  {\bibinfo {title} {{Two-season Atacama Cosmology Telescope polarimeter
  lensing power spectrum}},}\ }\href {\doibase 10.1103/PhysRevD.95.123529}
  {\bibfield  {journal} {\bibinfo  {journal} {Phys. Rev. D}\ }\textbf {\bibinfo
  {volume} {95}},\ \bibinfo {pages} {123529} (\bibinfo {year} {2017})},\
  \Eprint {http://arxiv.org/abs/1611.09753} {arXiv:1611.09753 [astro-ph.CO]}
  \BibitemShut {NoStop}%
\bibitem [{\citenamefont {Omori}\ \emph {et~al.}(2017)\citenamefont {Omori}
  \emph {et~al.}}]{Omori:2017tae}%
  \BibitemOpen
  \bibfield  {author} {\bibinfo {author} {\bibfnamefont {Y.}~\bibnamefont
  {Omori}} \emph {et~al.},\ }\bibfield  {title} {\enquote {\bibinfo {title} {{A
  2500 deg$^2$ CMB Lensing Map from Combined South Pole Telescope and Planck
  Data}},}\ }\href {\doibase 10.3847/1538-4357/aa8d1d} {\bibfield  {journal}
  {\bibinfo  {journal} {Astrophys. J.}\ }\textbf {\bibinfo {volume} {849}},\
  \bibinfo {pages} {124} (\bibinfo {year} {2017})},\ \Eprint
  {http://arxiv.org/abs/1705.00743} {arXiv:1705.00743 [astro-ph.CO]}
  \BibitemShut {NoStop}%
\bibitem [{\citenamefont {Aghanim}\ \emph
  {et~al.}(2020{\natexlab{a}})\citenamefont {Aghanim} \emph
  {et~al.}}]{Planck:2018lbu}%
  \BibitemOpen
  \bibfield  {author} {\bibinfo {author} {\bibfnamefont {N.}~\bibnamefont
  {Aghanim}} \emph {et~al.} (\bibinfo {collaboration} {Planck}),\ }\bibfield
  {title} {\enquote {\bibinfo {title} {{Planck 2018 results. VIII.
  Gravitational lensing}},}\ }\href {\doibase 10.1051/0004-6361/201833886}
  {\bibfield  {journal} {\bibinfo  {journal} {Astron. Astrophys.}\ }\textbf
  {\bibinfo {volume} {641}},\ \bibinfo {pages} {A8} (\bibinfo {year}
  {2020}{\natexlab{a}})},\ \Eprint {http://arxiv.org/abs/1807.06210}
  {arXiv:1807.06210 [astro-ph.CO]} \BibitemShut {NoStop}%
\bibitem [{\citenamefont {Wu}\ \emph {et~al.}(2019)\citenamefont {Wu} \emph
  {et~al.}}]{Wu:2019hek}%
  \BibitemOpen
  \bibfield  {author} {\bibinfo {author} {\bibfnamefont {W.~L.~K.}\
  \bibnamefont {Wu}} \emph {et~al.},\ }\bibfield  {title} {\enquote {\bibinfo
  {title} {{A Measurement of the Cosmic Microwave Background Lensing Potential
  and Power Spectrum from 500 deg$^2$ of SPTpol Temperature and Polarization
  Data}},}\ }\href {\doibase 10.3847/1538-4357/ab4186} {\bibfield  {journal}
  {\bibinfo  {journal} {Astrophys. J.}\ }\textbf {\bibinfo {volume} {884}},\
  \bibinfo {pages} {70} (\bibinfo {year} {2019})},\ \Eprint
  {http://arxiv.org/abs/1905.05777} {arXiv:1905.05777 [astro-ph.CO]}
  \BibitemShut {NoStop}%
\bibitem [{\citenamefont {Okamoto}\ and\ \citenamefont
  {Hu}(2003)}]{Okamoto:2003zw}%
  \BibitemOpen
  \bibfield  {author} {\bibinfo {author} {\bibfnamefont {Takemi}\ \bibnamefont
  {Okamoto}}\ and\ \bibinfo {author} {\bibfnamefont {Wayne}\ \bibnamefont
  {Hu}},\ }\bibfield  {title} {\enquote {\bibinfo {title} {{CMB lensing
  reconstruction on the full sky}},}\ }\href {\doibase
  10.1103/PhysRevD.67.083002} {\bibfield  {journal} {\bibinfo  {journal} {Phys.
  Rev.}\ }\textbf {\bibinfo {volume} {D67}},\ \bibinfo {pages} {083002}
  (\bibinfo {year} {2003})},\ \Eprint {http://arxiv.org/abs/astro-ph/0301031}
  {arXiv:astro-ph/0301031 [astro-ph]} \BibitemShut {NoStop}%
\bibitem [{\citenamefont {Abazajian}\ \emph {et~al.}(2016)\citenamefont
  {Abazajian} \emph {et~al.}}]{CMB-S4:2016ple}%
  \BibitemOpen
  \bibfield  {author} {\bibinfo {author} {\bibfnamefont {Kevork~N.}\
  \bibnamefont {Abazajian}} \emph {et~al.} (\bibinfo {collaboration}
  {CMB-S4}),\ }\bibfield  {title} {\enquote {\bibinfo {title} {{CMB-S4 Science
  Book, First Edition}},}\ }\href@noop {} {\  (\bibinfo {year} {2016})},\
  \Eprint {http://arxiv.org/abs/1610.02743} {arXiv:1610.02743 [astro-ph.CO]}
  \BibitemShut {NoStop}%
\bibitem [{\citenamefont {Hirata}\ and\ \citenamefont
  {Seljak}(2003{\natexlab{a}})}]{Hirata:2003ka}%
  \BibitemOpen
  \bibfield  {author} {\bibinfo {author} {\bibfnamefont {Christopher~M.}\
  \bibnamefont {Hirata}}\ and\ \bibinfo {author} {\bibfnamefont {Uros}\
  \bibnamefont {Seljak}},\ }\bibfield  {title} {\enquote {\bibinfo {title}
  {{Reconstruction of lensing from the cosmic microwave background
  polarization}},}\ }\href {\doibase 10.1103/PhysRevD.68.083002} {\bibfield
  {journal} {\bibinfo  {journal} {Phys. Rev.}\ }\textbf {\bibinfo {volume}
  {D68}},\ \bibinfo {pages} {083002} (\bibinfo {year} {2003}{\natexlab{a}})},\
  \Eprint {http://arxiv.org/abs/astro-ph/0306354} {arXiv:astro-ph/0306354
  [astro-ph]} \BibitemShut {NoStop}%
\bibitem [{\citenamefont {Hirata}\ and\ \citenamefont
  {Seljak}(2003{\natexlab{b}})}]{Hirata:2002jy}%
  \BibitemOpen
  \bibfield  {author} {\bibinfo {author} {\bibfnamefont {Christopher~M.}\
  \bibnamefont {Hirata}}\ and\ \bibinfo {author} {\bibfnamefont {Uros}\
  \bibnamefont {Seljak}},\ }\bibfield  {title} {\enquote {\bibinfo {title}
  {{Analyzing weak lensing of the cosmic microwave background using the
  likelihood function}},}\ }\href {\doibase 10.1103/PhysRevD.67.043001}
  {\bibfield  {journal} {\bibinfo  {journal} {Phys. Rev.}\ }\textbf {\bibinfo
  {volume} {D67}},\ \bibinfo {pages} {043001} (\bibinfo {year}
  {2003}{\natexlab{b}})},\ \Eprint {http://arxiv.org/abs/astro-ph/0209489}
  {arXiv:astro-ph/0209489 [astro-ph]} \BibitemShut {NoStop}%
\bibitem [{\citenamefont {Millea}\ \emph {et~al.}(2019)\citenamefont {Millea},
  \citenamefont {Anderes},\ and\ \citenamefont {Wandelt}}]{Millea:2017fyd}%
  \BibitemOpen
  \bibfield  {author} {\bibinfo {author} {\bibfnamefont {Marius}\ \bibnamefont
  {Millea}}, \bibinfo {author} {\bibfnamefont {Ethan}\ \bibnamefont {Anderes}},
  \ and\ \bibinfo {author} {\bibfnamefont {Benjamin~D.}\ \bibnamefont
  {Wandelt}},\ }\bibfield  {title} {\enquote {\bibinfo {title} {{Bayesian
  delensing of CMB temperature and polarization}},}\ }\href {\doibase
  10.1103/PhysRevD.100.023509} {\bibfield  {journal} {\bibinfo  {journal}
  {Phys. Rev. D}\ }\textbf {\bibinfo {volume} {100}},\ \bibinfo {pages}
  {023509} (\bibinfo {year} {2019})},\ \Eprint
  {http://arxiv.org/abs/1708.06753} {arXiv:1708.06753 [astro-ph.CO]}
  \BibitemShut {NoStop}%
\bibitem [{\citenamefont {Millea}\ \emph {et~al.}(2020)\citenamefont {Millea},
  \citenamefont {Anderes},\ and\ \citenamefont {Wandelt}}]{Millea:2020cpw}%
  \BibitemOpen
  \bibfield  {author} {\bibinfo {author} {\bibfnamefont {Marius}\ \bibnamefont
  {Millea}}, \bibinfo {author} {\bibfnamefont {Ethan}\ \bibnamefont {Anderes}},
  \ and\ \bibinfo {author} {\bibfnamefont {Benjamin~D.}\ \bibnamefont
  {Wandelt}},\ }\bibfield  {title} {\enquote {\bibinfo {title} {{Sampling-based
  inference of the primordial CMB and gravitational lensing}},}\ }\href
  {\doibase 10.1103/PhysRevD.102.123542} {\bibfield  {journal} {\bibinfo
  {journal} {Phys. Rev. D}\ }\textbf {\bibinfo {volume} {102}},\ \bibinfo
  {pages} {123542} (\bibinfo {year} {2020})},\ \Eprint
  {http://arxiv.org/abs/2002.00965} {arXiv:2002.00965 [astro-ph.CO]}
  \BibitemShut {NoStop}%
\bibitem [{\citenamefont {Carron}\ and\ \citenamefont
  {Lewis}(2017)}]{Carron:2017mqf}%
  \BibitemOpen
  \bibfield  {author} {\bibinfo {author} {\bibfnamefont {Julien}\ \bibnamefont
  {Carron}}\ and\ \bibinfo {author} {\bibfnamefont {Antony}\ \bibnamefont
  {Lewis}},\ }\bibfield  {title} {\enquote {\bibinfo {title} {{Maximum a
  posteriori CMB lensing reconstruction}},}\ }\href {\doibase
  10.1103/PhysRevD.96.063510} {\bibfield  {journal} {\bibinfo  {journal} {Phys.
  Rev. D}\ }\textbf {\bibinfo {volume} {96}},\ \bibinfo {pages} {063510}
  (\bibinfo {year} {2017})},\ \Eprint {http://arxiv.org/abs/1704.08230}
  {arXiv:1704.08230 [astro-ph.CO]} \BibitemShut {NoStop}%
\bibitem [{\citenamefont {Carron}(2019)}]{Carron:2018lcr}%
  \BibitemOpen
  \bibfield  {author} {\bibinfo {author} {\bibfnamefont {Julien}\ \bibnamefont
  {Carron}},\ }\bibfield  {title} {\enquote {\bibinfo {title} {{Optimal
  constraints on primordial gravitational waves from the lensed CMB}},}\ }\href
  {\doibase 10.1103/PhysRevD.99.043518} {\bibfield  {journal} {\bibinfo
  {journal} {Phys. Rev. D}\ }\textbf {\bibinfo {volume} {99}},\ \bibinfo
  {pages} {043518} (\bibinfo {year} {2019})},\ \Eprint
  {http://arxiv.org/abs/1808.10349} {arXiv:1808.10349 [astro-ph.CO]}
  \BibitemShut {NoStop}%
\bibitem [{\citenamefont {Hu}\ and\ \citenamefont {Okamoto}(2002)}]{Hu:2001kj}%
  \BibitemOpen
  \bibfield  {author} {\bibinfo {author} {\bibfnamefont {Wayne}\ \bibnamefont
  {Hu}}\ and\ \bibinfo {author} {\bibfnamefont {Takemi}\ \bibnamefont
  {Okamoto}},\ }\bibfield  {title} {\enquote {\bibinfo {title} {{Mass
  reconstruction with cmb polarization}},}\ }\href {\doibase 10.1086/341110}
  {\bibfield  {journal} {\bibinfo  {journal} {Astrophys. J.}\ }\textbf
  {\bibinfo {volume} {574}},\ \bibinfo {pages} {566--574} (\bibinfo {year}
  {2002})},\ \Eprint {http://arxiv.org/abs/astro-ph/0111606}
  {arXiv:astro-ph/0111606} \BibitemShut {NoStop}%
\bibitem [{\citenamefont {Kesden}\ \emph {et~al.}(2003)\citenamefont {Kesden},
  \citenamefont {Cooray},\ and\ \citenamefont {Kamionkowski}}]{Kesden:2003cc}%
  \BibitemOpen
  \bibfield  {author} {\bibinfo {author} {\bibfnamefont {Michael~H.}\
  \bibnamefont {Kesden}}, \bibinfo {author} {\bibfnamefont {Asantha}\
  \bibnamefont {Cooray}}, \ and\ \bibinfo {author} {\bibfnamefont {Marc}\
  \bibnamefont {Kamionkowski}},\ }\bibfield  {title} {\enquote {\bibinfo
  {title} {{Lensing reconstruction with CMB temperature and polarization}},}\
  }\href {\doibase 10.1103/PhysRevD.67.123507} {\bibfield  {journal} {\bibinfo
  {journal} {Phys. Rev. D}\ }\textbf {\bibinfo {volume} {67}},\ \bibinfo
  {pages} {123507} (\bibinfo {year} {2003})},\ \Eprint
  {http://arxiv.org/abs/astro-ph/0302536} {arXiv:astro-ph/0302536} \BibitemShut
  {NoStop}%
\bibitem [{\citenamefont {Hanson}\ \emph {et~al.}(2011)\citenamefont {Hanson},
  \citenamefont {Challinor}, \citenamefont {Efstathiou},\ and\ \citenamefont
  {Bielewicz}}]{Hanson:2010rp}%
  \BibitemOpen
  \bibfield  {author} {\bibinfo {author} {\bibfnamefont {Duncan}\ \bibnamefont
  {Hanson}}, \bibinfo {author} {\bibfnamefont {Anthony}\ \bibnamefont
  {Challinor}}, \bibinfo {author} {\bibfnamefont {George}\ \bibnamefont
  {Efstathiou}}, \ and\ \bibinfo {author} {\bibfnamefont {Pawel}\ \bibnamefont
  {Bielewicz}},\ }\bibfield  {title} {\enquote {\bibinfo {title} {{CMB
  temperature lensing power reconstruction}},}\ }\href {\doibase
  10.1103/PhysRevD.83.043005} {\bibfield  {journal} {\bibinfo  {journal} {Phys.
  Rev. D}\ }\textbf {\bibinfo {volume} {83}},\ \bibinfo {pages} {043005}
  (\bibinfo {year} {2011})},\ \Eprint {http://arxiv.org/abs/1008.4403}
  {arXiv:1008.4403 [astro-ph.CO]} \BibitemShut {NoStop}%
\bibitem [{\citenamefont {B{\"o}hm}\ \emph {et~al.}(2016)\citenamefont
  {B{\"o}hm}, \citenamefont {Schmittfull},\ and\ \citenamefont
  {Sherwin}}]{Bohm:2016gzt}%
  \BibitemOpen
  \bibfield  {author} {\bibinfo {author} {\bibfnamefont {Vanessa}\ \bibnamefont
  {B{\"o}hm}}, \bibinfo {author} {\bibfnamefont {Marcel}\ \bibnamefont
  {Schmittfull}}, \ and\ \bibinfo {author} {\bibfnamefont {Blake~D.}\
  \bibnamefont {Sherwin}},\ }\bibfield  {title} {\enquote {\bibinfo {title}
  {{Bias to CMB lensing measurements from the bispectrum of large-scale
  structure}},}\ }\href {\doibase 10.1103/PhysRevD.94.043519} {\bibfield
  {journal} {\bibinfo  {journal} {Phys. Rev.}\ }\textbf {\bibinfo {volume}
  {D94}},\ \bibinfo {pages} {043519} (\bibinfo {year} {2016})},\ \Eprint
  {http://arxiv.org/abs/1605.01392} {arXiv:1605.01392 [astro-ph.CO]}
  \BibitemShut {NoStop}%
\bibitem [{\citenamefont {B\"ohm}\ \emph {et~al.}(2018)\citenamefont {B\"ohm},
  \citenamefont {Sherwin}, \citenamefont {Liu}, \citenamefont {Hill},
  \citenamefont {Schmittfull},\ and\ \citenamefont {Namikawa}}]{Bohm:2018omn}%
  \BibitemOpen
  \bibfield  {author} {\bibinfo {author} {\bibfnamefont {Vanessa}\ \bibnamefont
  {B\"ohm}}, \bibinfo {author} {\bibfnamefont {Blake~D.}\ \bibnamefont
  {Sherwin}}, \bibinfo {author} {\bibfnamefont {Jia}\ \bibnamefont {Liu}},
  \bibinfo {author} {\bibfnamefont {J.~Colin}\ \bibnamefont {Hill}}, \bibinfo
  {author} {\bibfnamefont {Marcel}\ \bibnamefont {Schmittfull}}, \ and\
  \bibinfo {author} {\bibfnamefont {Toshiya}\ \bibnamefont {Namikawa}},\
  }\bibfield  {title} {\enquote {\bibinfo {title} {{Effect of non-Gaussian
  lensing deflections on CMB lensing measurements}},}\ }\href {\doibase
  10.1103/PhysRevD.98.123510} {\bibfield  {journal} {\bibinfo  {journal} {Phys.
  Rev. D}\ }\textbf {\bibinfo {volume} {98}},\ \bibinfo {pages} {123510}
  (\bibinfo {year} {2018})},\ \Eprint {http://arxiv.org/abs/1806.01157}
  {arXiv:1806.01157 [astro-ph.CO]} \BibitemShut {NoStop}%
\bibitem [{\citenamefont {Beck}\ \emph {et~al.}(2018)\citenamefont {Beck},
  \citenamefont {Fabbian},\ and\ \citenamefont {Errard}}]{Beck:2018wud}%
  \BibitemOpen
  \bibfield  {author} {\bibinfo {author} {\bibfnamefont {Dominic}\ \bibnamefont
  {Beck}}, \bibinfo {author} {\bibfnamefont {Giulio}\ \bibnamefont {Fabbian}},
  \ and\ \bibinfo {author} {\bibfnamefont {Josquin}\ \bibnamefont {Errard}},\
  }\bibfield  {title} {\enquote {\bibinfo {title} {{Lensing Reconstruction in
  Post-Born Cosmic Microwave Background Weak Lensing}},}\ }\href {\doibase
  10.1103/PhysRevD.98.043512} {\bibfield  {journal} {\bibinfo  {journal} {Phys.
  Rev. D}\ }\textbf {\bibinfo {volume} {98}},\ \bibinfo {pages} {043512}
  (\bibinfo {year} {2018})},\ \Eprint {http://arxiv.org/abs/1806.01216}
  {arXiv:1806.01216 [astro-ph.CO]} \BibitemShut {NoStop}%
\bibitem [{\citenamefont {Fabbian}\ \emph {et~al.}(2019)\citenamefont
  {Fabbian}, \citenamefont {Lewis},\ and\ \citenamefont
  {Beck}}]{Fabbian:2019tik}%
  \BibitemOpen
  \bibfield  {author} {\bibinfo {author} {\bibfnamefont {Giulio}\ \bibnamefont
  {Fabbian}}, \bibinfo {author} {\bibfnamefont {Antony}\ \bibnamefont {Lewis}},
  \ and\ \bibinfo {author} {\bibfnamefont {Dominic}\ \bibnamefont {Beck}},\
  }\bibfield  {title} {\enquote {\bibinfo {title} {{CMB lensing reconstruction
  biases in cross-correlation with large-scale structure probes}},}\ }\href
  {\doibase 10.1088/1475-7516/2019/10/057} {\bibfield  {journal} {\bibinfo
  {journal} {JCAP}\ }\textbf {\bibinfo {volume} {10}},\ \bibinfo {pages} {057}
  (\bibinfo {year} {2019})},\ \Eprint {http://arxiv.org/abs/1906.08760}
  {arXiv:1906.08760 [astro-ph.CO]} \BibitemShut {NoStop}%
\bibitem [{\citenamefont {Millea}\ \emph {et~al.}(2021)\citenamefont {Millea}
  \emph {et~al.}}]{Millea:2020iuw}%
  \BibitemOpen
  \bibfield  {author} {\bibinfo {author} {\bibfnamefont {M.}~\bibnamefont
  {Millea}} \emph {et~al.},\ }\bibfield  {title} {\enquote {\bibinfo {title}
  {{Optimal Cosmic Microwave Background Lensing Reconstruction and Parameter
  Estimation with SPTpol Data}},}\ }\href {\doibase 10.3847/1538-4357/ac02bb}
  {\bibfield  {journal} {\bibinfo  {journal} {Astrophys. J.}\ }\textbf
  {\bibinfo {volume} {922}},\ \bibinfo {pages} {259} (\bibinfo {year}
  {2021})},\ \Eprint {http://arxiv.org/abs/2012.01709} {arXiv:2012.01709
  [astro-ph.CO]} \BibitemShut {NoStop}%
\bibitem [{\citenamefont {collaboration}(in prep.)}]{CMBS4:inprep}%
  \BibitemOpen
  \bibfield  {author} {\bibinfo {author} {\bibfnamefont {The CMB-S4}\
  \bibnamefont {collaboration}},\ }\bibfield  {title} {\enquote {\bibinfo
  {title} {{CMB-S4: Iterative internal delensing and $r$ constraints}},}\
  }\href@noop {} {\  (\bibinfo {year} {in prep.})}\BibitemShut {NoStop}%
\bibitem [{\citenamefont {Namikawa}\ \emph {et~al.}(2013)\citenamefont
  {Namikawa}, \citenamefont {Hanson},\ and\ \citenamefont
  {Takahashi}}]{Namikawa:2012pe}%
  \BibitemOpen
  \bibfield  {author} {\bibinfo {author} {\bibfnamefont {Toshiya}\ \bibnamefont
  {Namikawa}}, \bibinfo {author} {\bibfnamefont {Duncan}\ \bibnamefont
  {Hanson}}, \ and\ \bibinfo {author} {\bibfnamefont {Ryuichi}\ \bibnamefont
  {Takahashi}},\ }\bibfield  {title} {\enquote {\bibinfo {title}
  {{Bias-Hardened CMB Lensing}},}\ }\href {\doibase 10.1093/mnras/stt195}
  {\bibfield  {journal} {\bibinfo  {journal} {Mon. Not. Roy. Astron. Soc.}\
  }\textbf {\bibinfo {volume} {431}},\ \bibinfo {pages} {609--620} (\bibinfo
  {year} {2013})},\ \Eprint {http://arxiv.org/abs/1209.0091} {arXiv:1209.0091
  [astro-ph.CO]} \BibitemShut {NoStop}%
\bibitem [{\citenamefont {Story}\ \emph {et~al.}(2015)\citenamefont {Story}
  \emph {et~al.}}]{Story:2014hni}%
  \BibitemOpen
  \bibfield  {author} {\bibinfo {author} {\bibfnamefont {K.~T.}\ \bibnamefont
  {Story}} \emph {et~al.} (\bibinfo {collaboration} {SPT}),\ }\bibfield
  {title} {\enquote {\bibinfo {title} {{A Measurement of the Cosmic Microwave
  Background Gravitational Lensing Potential from 100 Square Degrees of SPTpol
  Data}},}\ }\href {\doibase 10.1088/0004-637X/810/1/50} {\bibfield  {journal}
  {\bibinfo  {journal} {Astrophys. J.}\ }\textbf {\bibinfo {volume} {810}},\
  \bibinfo {pages} {50} (\bibinfo {year} {2015})},\ \Eprint
  {http://arxiv.org/abs/1412.4760} {arXiv:1412.4760 [astro-ph.CO]} \BibitemShut
  {NoStop}%
\bibitem [{\citenamefont {Ade}\ \emph {et~al.}(2016{\natexlab{a}})\citenamefont
  {Ade} \emph {et~al.}}]{Ade:2015zua}%
  \BibitemOpen
  \bibfield  {author} {\bibinfo {author} {\bibfnamefont {P.~A.~R.}\
  \bibnamefont {Ade}} \emph {et~al.} (\bibinfo {collaboration} {Planck}),\
  }\bibfield  {title} {\enquote {\bibinfo {title} {{Planck 2015 results. XV.
  Gravitational lensing}},}\ }\href {\doibase 10.1051/0004-6361/201525941}
  {\bibfield  {journal} {\bibinfo  {journal} {Astron. Astrophys.}\ }\textbf
  {\bibinfo {volume} {594}},\ \bibinfo {pages} {A15} (\bibinfo {year}
  {2016}{\natexlab{a}})},\ \Eprint {http://arxiv.org/abs/1502.01591}
  {arXiv:1502.01591 [astro-ph.CO]} \BibitemShut {NoStop}%
\bibitem [{\citenamefont {Peloton}\ \emph {et~al.}(2017)\citenamefont
  {Peloton}, \citenamefont {Schmittfull}, \citenamefont {Lewis}, \citenamefont
  {Carron},\ and\ \citenamefont {Zahn}}]{Peloton:2016kbw}%
  \BibitemOpen
  \bibfield  {author} {\bibinfo {author} {\bibfnamefont {Julien}\ \bibnamefont
  {Peloton}}, \bibinfo {author} {\bibfnamefont {Marcel}\ \bibnamefont
  {Schmittfull}}, \bibinfo {author} {\bibfnamefont {Antony}\ \bibnamefont
  {Lewis}}, \bibinfo {author} {\bibfnamefont {Julien}\ \bibnamefont {Carron}},
  \ and\ \bibinfo {author} {\bibfnamefont {Oliver}\ \bibnamefont {Zahn}},\
  }\bibfield  {title} {\enquote {\bibinfo {title} {{Full covariance of CMB and
  lensing reconstruction power spectra}},}\ }\href {\doibase
  10.1103/PhysRevD.95.043508} {\bibfield  {journal} {\bibinfo  {journal} {Phys.
  Rev.}\ }\textbf {\bibinfo {volume} {D95}},\ \bibinfo {pages} {043508}
  (\bibinfo {year} {2017})},\ \Eprint {http://arxiv.org/abs/1611.01446}
  {arXiv:1611.01446 [astro-ph.CO]} \BibitemShut {NoStop}%
\bibitem [{\citenamefont {Adachi}\ \emph {et~al.}(2020)\citenamefont {Adachi}
  \emph {et~al.}}]{POLARBEAR:2019snn}%
  \BibitemOpen
  \bibfield  {author} {\bibinfo {author} {\bibfnamefont {S.}~\bibnamefont
  {Adachi}} \emph {et~al.} (\bibinfo {collaboration} {POLARBEAR}),\ }\bibfield
  {title} {\enquote {\bibinfo {title} {{Internal delensing of Cosmic Microwave
  Background polarization $B$-modes with the POLARBEAR experiment}},}\ }\href
  {\doibase 10.1103/PhysRevLett.124.131301} {\bibfield  {journal} {\bibinfo
  {journal} {Phys. Rev. Lett.}\ }\textbf {\bibinfo {volume} {124}},\ \bibinfo
  {pages} {131301} (\bibinfo {year} {2020})},\ \Eprint
  {http://arxiv.org/abs/1909.13832} {arXiv:1909.13832 [astro-ph.CO]}
  \BibitemShut {NoStop}%
\bibitem [{\citenamefont {Maniyar}\ \emph {et~al.}(2021)\citenamefont
  {Maniyar}, \citenamefont {Ali-Ha\"\i{}moud}, \citenamefont {Carron},
  \citenamefont {Lewis},\ and\ \citenamefont
  {Madhavacheril}}]{Maniyar:2021msb}%
  \BibitemOpen
  \bibfield  {author} {\bibinfo {author} {\bibfnamefont {Abhishek~S.}\
  \bibnamefont {Maniyar}}, \bibinfo {author} {\bibfnamefont {Yacine}\
  \bibnamefont {Ali-Ha\"\i{}moud}}, \bibinfo {author} {\bibfnamefont {Julien}\
  \bibnamefont {Carron}}, \bibinfo {author} {\bibfnamefont {Antony}\
  \bibnamefont {Lewis}}, \ and\ \bibinfo {author} {\bibfnamefont {Mathew~S.}\
  \bibnamefont {Madhavacheril}},\ }\bibfield  {title} {\enquote {\bibinfo
  {title} {{Quadratic estimators for CMB weak lensing}},}\ }\href {\doibase
  10.1103/PhysRevD.103.083524} {\bibfield  {journal} {\bibinfo  {journal}
  {Phys. Rev. D}\ }\textbf {\bibinfo {volume} {103}},\ \bibinfo {pages}
  {083524} (\bibinfo {year} {2021})},\ \Eprint
  {http://arxiv.org/abs/2101.12193} {arXiv:2101.12193 [astro-ph.CO]}
  \BibitemShut {NoStop}%
\bibitem [{\citenamefont {Lewis}\ \emph {et~al.}(2011)\citenamefont {Lewis},
  \citenamefont {Challinor},\ and\ \citenamefont {Hanson}}]{Lewis:2011fk}%
  \BibitemOpen
  \bibfield  {author} {\bibinfo {author} {\bibfnamefont {Antony}\ \bibnamefont
  {Lewis}}, \bibinfo {author} {\bibfnamefont {Anthony}\ \bibnamefont
  {Challinor}}, \ and\ \bibinfo {author} {\bibfnamefont {Duncan}\ \bibnamefont
  {Hanson}},\ }\bibfield  {title} {\enquote {\bibinfo {title} {{The shape of
  the CMB lensing bispectrum}},}\ }\href {\doibase
  10.1088/1475-7516/2011/03/018} {\bibfield  {journal} {\bibinfo  {journal}
  {JCAP}\ }\textbf {\bibinfo {volume} {03}},\ \bibinfo {pages} {018} (\bibinfo
  {year} {2011})},\ \Eprint {http://arxiv.org/abs/1101.2234} {arXiv:1101.2234
  [astro-ph.CO]} \BibitemShut {NoStop}%
\bibitem [{\citenamefont {Smith}\ \emph {et~al.}(2012)\citenamefont {Smith},
  \citenamefont {Hanson}, \citenamefont {LoVerde}, \citenamefont {Hirata},\
  and\ \citenamefont {Zahn}}]{Smith:2010gu}%
  \BibitemOpen
  \bibfield  {author} {\bibinfo {author} {\bibfnamefont {Kendrick~M.}\
  \bibnamefont {Smith}}, \bibinfo {author} {\bibfnamefont {Duncan}\
  \bibnamefont {Hanson}}, \bibinfo {author} {\bibfnamefont {Marilena}\
  \bibnamefont {LoVerde}}, \bibinfo {author} {\bibfnamefont {Christopher~M.}\
  \bibnamefont {Hirata}}, \ and\ \bibinfo {author} {\bibfnamefont {Oliver}\
  \bibnamefont {Zahn}},\ }\bibfield  {title} {\enquote {\bibinfo {title}
  {{Delensing CMB Polarization with External Datasets}},}\ }\href {\doibase
  10.1088/1475-7516/2012/06/014} {\bibfield  {journal} {\bibinfo  {journal}
  {JCAP}\ }\textbf {\bibinfo {volume} {06}},\ \bibinfo {pages} {014} (\bibinfo
  {year} {2012})},\ \Eprint {http://arxiv.org/abs/1010.0048} {arXiv:1010.0048
  [astro-ph.CO]} \BibitemShut {NoStop}%
\bibitem [{\citenamefont {Hotinli}\ \emph {et~al.}(2021)\citenamefont
  {Hotinli}, \citenamefont {Meyers}, \citenamefont {Trendafilova},
  \citenamefont {Green},\ and\ \citenamefont {van Engelen}}]{Hotinli:2021umk}%
  \BibitemOpen
  \bibfield  {author} {\bibinfo {author} {\bibfnamefont {Selim~C.}\
  \bibnamefont {Hotinli}}, \bibinfo {author} {\bibfnamefont {Joel}\
  \bibnamefont {Meyers}}, \bibinfo {author} {\bibfnamefont {Cynthia}\
  \bibnamefont {Trendafilova}}, \bibinfo {author} {\bibfnamefont {Daniel}\
  \bibnamefont {Green}}, \ and\ \bibinfo {author} {\bibfnamefont {Alexander}\
  \bibnamefont {van Engelen}},\ }\bibfield  {title} {\enquote {\bibinfo {title}
  {{The Benefits of CMB Delensing}},}\ }\href@noop {} {\  (\bibinfo {year}
  {2021})},\ \Eprint {http://arxiv.org/abs/2111.15036} {arXiv:2111.15036
  [astro-ph.CO]} \BibitemShut {NoStop}%
\bibitem [{\citenamefont {Simard}\ \emph {et~al.}(2017)\citenamefont {Simard}
  \emph {et~al.}}]{Simard:2017xtw}%
  \BibitemOpen
  \bibfield  {author} {\bibinfo {author} {\bibfnamefont {G.}~\bibnamefont
  {Simard}} \emph {et~al.} (\bibinfo {collaboration} {SPT}),\ }\bibfield
  {title} {\enquote {\bibinfo {title} {{Constraints on Cosmological Parameters
  from the Angular Power Spectrum of a Combined 2500 deg$^2$ SPT-SZ and Planck
  Gravitational Lensing Map}},}\ }\href@noop {} {\bibfield  {journal} {\bibinfo
   {journal} {Submitted to: Astrophys. J.}\ } (\bibinfo {year} {2017})},\
  \Eprint {http://arxiv.org/abs/1712.07541} {arXiv:1712.07541 [astro-ph.CO]}
  \BibitemShut {NoStop}%
\bibitem [{\citenamefont {Torrado}\ and\ \citenamefont
  {Lewis}(2021)}]{Torrado:2020dgo}%
  \BibitemOpen
  \bibfield  {author} {\bibinfo {author} {\bibfnamefont {Jesus}\ \bibnamefont
  {Torrado}}\ and\ \bibinfo {author} {\bibfnamefont {Antony}\ \bibnamefont
  {Lewis}},\ }\bibfield  {title} {\enquote {\bibinfo {title} {{Cobaya: Code for
  Bayesian Analysis of hierarchical physical models}},}\ }\href {\doibase
  10.1088/1475-7516/2021/05/057} {\bibfield  {journal} {\bibinfo  {journal}
  {JCAP}\ }\textbf {\bibinfo {volume} {05}},\ \bibinfo {pages} {057} (\bibinfo
  {year} {2021})},\ \Eprint {http://arxiv.org/abs/2005.05290} {arXiv:2005.05290
  [astro-ph.IM]} \BibitemShut {NoStop}%
\bibitem [{\citenamefont {Lewis}\ and\ \citenamefont
  {Bridle}(2002)}]{Lewis:2002ah}%
  \BibitemOpen
  \bibfield  {author} {\bibinfo {author} {\bibfnamefont {Antony}\ \bibnamefont
  {Lewis}}\ and\ \bibinfo {author} {\bibfnamefont {Sarah}\ \bibnamefont
  {Bridle}},\ }\bibfield  {title} {\enquote {\bibinfo {title} {{Cosmological
  parameters from CMB and other data: A Monte Carlo approach}},}\ }\href
  {\doibase 10.1103/PhysRevD.66.103511} {\bibfield  {journal} {\bibinfo
  {journal} {Phys. Rev.}\ }\textbf {\bibinfo {volume} {D66}},\ \bibinfo {pages}
  {103511} (\bibinfo {year} {2002})},\ \Eprint
  {http://arxiv.org/abs/astro-ph/0205436} {arXiv:astro-ph/0205436 [astro-ph]}
  \BibitemShut {NoStop}%
\bibitem [{\citenamefont {Lewis}(2013)}]{Lewis:2013hha}%
  \BibitemOpen
  \bibfield  {author} {\bibinfo {author} {\bibfnamefont {Antony}\ \bibnamefont
  {Lewis}},\ }\bibfield  {title} {\enquote {\bibinfo {title} {{Efficient
  sampling of fast and slow cosmological parameters}},}\ }\href {\doibase
  10.1103/PhysRevD.87.103529} {\bibfield  {journal} {\bibinfo  {journal} {Phys.
  Rev.}\ }\textbf {\bibinfo {volume} {D87}},\ \bibinfo {pages} {103529}
  (\bibinfo {year} {2013})},\ \Eprint {http://arxiv.org/abs/1304.4473}
  {arXiv:1304.4473 [astro-ph.CO]} \BibitemShut {NoStop}%
\bibitem [{\citenamefont {Lewis}\ \emph {et~al.}(2000)\citenamefont {Lewis},
  \citenamefont {Challinor},\ and\ \citenamefont {Lasenby}}]{Lewis:1999bs}%
  \BibitemOpen
  \bibfield  {author} {\bibinfo {author} {\bibfnamefont {Antony}\ \bibnamefont
  {Lewis}}, \bibinfo {author} {\bibfnamefont {Anthony}\ \bibnamefont
  {Challinor}}, \ and\ \bibinfo {author} {\bibfnamefont {Anthony}\ \bibnamefont
  {Lasenby}},\ }\bibfield  {title} {\enquote {\bibinfo {title} {{Efficient
  computation of CMB anisotropies in closed FRW models}},}\ }\href {\doibase
  10.1086/309179} {\bibfield  {journal} {\bibinfo  {journal} {Astrophys. J.}\
  }\textbf {\bibinfo {volume} {538}},\ \bibinfo {pages} {473--476} (\bibinfo
  {year} {2000})},\ \Eprint {http://arxiv.org/abs/astro-ph/9911177}
  {arXiv:astro-ph/9911177 [astro-ph]} \BibitemShut {NoStop}%
\bibitem [{\citenamefont {Howlett}\ \emph {et~al.}(2012)\citenamefont
  {Howlett}, \citenamefont {Lewis}, \citenamefont {Hall},\ and\ \citenamefont
  {Challinor}}]{Howlett:2012mh}%
  \BibitemOpen
  \bibfield  {author} {\bibinfo {author} {\bibfnamefont {Cullan}\ \bibnamefont
  {Howlett}}, \bibinfo {author} {\bibfnamefont {Antony}\ \bibnamefont {Lewis}},
  \bibinfo {author} {\bibfnamefont {Alex}\ \bibnamefont {Hall}}, \ and\
  \bibinfo {author} {\bibfnamefont {Anthony}\ \bibnamefont {Challinor}},\
  }\bibfield  {title} {\enquote {\bibinfo {title} {{CMB power spectrum
  parameter degeneracies in the era of precision cosmology}},}\ }\href
  {\doibase 10.1088/1475-7516/2012/04/027} {\bibfield  {journal} {\bibinfo
  {journal} {JCAP}\ }\textbf {\bibinfo {volume} {1204}},\ \bibinfo {pages}
  {027} (\bibinfo {year} {2012})},\ \Eprint {http://arxiv.org/abs/1201.3654}
  {arXiv:1201.3654 [astro-ph.CO]} \BibitemShut {NoStop}%
\bibitem [{\citenamefont {Pan}\ \emph {et~al.}(2014)\citenamefont {Pan},
  \citenamefont {Knox},\ and\ \citenamefont {White}}]{Pan:2014xua}%
  \BibitemOpen
  \bibfield  {author} {\bibinfo {author} {\bibfnamefont {Z.}~\bibnamefont
  {Pan}}, \bibinfo {author} {\bibfnamefont {L.}~\bibnamefont {Knox}}, \ and\
  \bibinfo {author} {\bibfnamefont {M.}~\bibnamefont {White}},\ }\bibfield
  {title} {\enquote {\bibinfo {title} {{Dependence of the Cosmic Microwave
  Background Lensing Power Spectrum on the Matter Density}},}\ }\href {\doibase
  10.1093/mnras/stu1971} {\bibfield  {journal} {\bibinfo  {journal} {Mon. Not.
  Roy. Astron. Soc.}\ }\textbf {\bibinfo {volume} {445}},\ \bibinfo {pages}
  {2941--2945} (\bibinfo {year} {2014})},\ \Eprint
  {http://arxiv.org/abs/1406.5459} {arXiv:1406.5459 [astro-ph.CO]} \BibitemShut
  {NoStop}%
\bibitem [{\citenamefont {Ade}\ \emph {et~al.}(2016{\natexlab{b}})\citenamefont
  {Ade} \emph {et~al.}}]{Planck:2015mym}%
  \BibitemOpen
  \bibfield  {author} {\bibinfo {author} {\bibfnamefont {P.~A.~R.}\
  \bibnamefont {Ade}} \emph {et~al.} (\bibinfo {collaboration} {Planck}),\
  }\bibfield  {title} {\enquote {\bibinfo {title} {{Planck 2015 results. XV.
  Gravitational lensing}},}\ }\href {\doibase 10.1051/0004-6361/201525941}
  {\bibfield  {journal} {\bibinfo  {journal} {Astron. Astrophys.}\ }\textbf
  {\bibinfo {volume} {594}},\ \bibinfo {pages} {A15} (\bibinfo {year}
  {2016}{\natexlab{b}})},\ \Eprint {http://arxiv.org/abs/1502.01591}
  {arXiv:1502.01591 [astro-ph.CO]} \BibitemShut {NoStop}%
\bibitem [{\citenamefont {Bianchini}\ \emph {et~al.}(2020)\citenamefont
  {Bianchini} \emph {et~al.}}]{SPT:2019fqo}%
  \BibitemOpen
  \bibfield  {author} {\bibinfo {author} {\bibfnamefont {F.}~\bibnamefont
  {Bianchini}} \emph {et~al.} (\bibinfo {collaboration} {SPT}),\ }\bibfield
  {title} {\enquote {\bibinfo {title} {{Constraints on Cosmological Parameters
  from the 500 deg$^2$ SPTpol Lensing Power Spectrum}},}\ }\href {\doibase
  10.3847/1538-4357/ab6082} {\bibfield  {journal} {\bibinfo  {journal}
  {Astrophys. J.}\ }\textbf {\bibinfo {volume} {888}},\ \bibinfo {pages} {119}
  (\bibinfo {year} {2020})},\ \Eprint {http://arxiv.org/abs/1910.07157}
  {arXiv:1910.07157 [astro-ph.CO]} \BibitemShut {NoStop}%
\bibitem [{\citenamefont {Lesgourgues}\ and\ \citenamefont
  {Pastor}(2006)}]{Lesgourgues:2006nd}%
  \BibitemOpen
  \bibfield  {author} {\bibinfo {author} {\bibfnamefont {Julien}\ \bibnamefont
  {Lesgourgues}}\ and\ \bibinfo {author} {\bibfnamefont {Sergio}\ \bibnamefont
  {Pastor}},\ }\bibfield  {title} {\enquote {\bibinfo {title} {{Massive
  neutrinos and cosmology}},}\ }\href {\doibase 10.1016/j.physrep.2006.04.001}
  {\bibfield  {journal} {\bibinfo  {journal} {Phys. Rept.}\ }\textbf {\bibinfo
  {volume} {429}},\ \bibinfo {pages} {307--379} (\bibinfo {year} {2006})},\
  \Eprint {http://arxiv.org/abs/astro-ph/0603494} {arXiv:astro-ph/0603494}
  \BibitemShut {NoStop}%
\bibitem [{\citenamefont {Percival}\ \emph {et~al.}(2002)\citenamefont
  {Percival} \emph {et~al.}}]{2dFGRSTeam:2002tzq}%
  \BibitemOpen
  \bibfield  {author} {\bibinfo {author} {\bibfnamefont {Will~J.}\ \bibnamefont
  {Percival}} \emph {et~al.} (\bibinfo {collaboration} {2dFGRS Team}),\
  }\bibfield  {title} {\enquote {\bibinfo {title} {{Parameter constraints for
  flat cosmologies from CMB and 2dFGRS power spectra}},}\ }\href {\doibase
  10.1046/j.1365-8711.2002.06001.x} {\bibfield  {journal} {\bibinfo  {journal}
  {Mon. Not. Roy. Astron. Soc.}\ }\textbf {\bibinfo {volume} {337}},\ \bibinfo
  {pages} {1068} (\bibinfo {year} {2002})},\ \Eprint
  {http://arxiv.org/abs/astro-ph/0206256} {arXiv:astro-ph/0206256} \BibitemShut
  {NoStop}%
\bibitem [{\citenamefont {Aghanim}\ \emph
  {et~al.}(2020{\natexlab{b}})\citenamefont {Aghanim} \emph
  {et~al.}}]{Planck:2018vyg}%
  \BibitemOpen
  \bibfield  {author} {\bibinfo {author} {\bibfnamefont {N.}~\bibnamefont
  {Aghanim}} \emph {et~al.} (\bibinfo {collaboration} {Planck}),\ }\bibfield
  {title} {\enquote {\bibinfo {title} {{Planck 2018 results. VI. Cosmological
  parameters}},}\ }\href {\doibase 10.1051/0004-6361/201833910} {\bibfield
  {journal} {\bibinfo  {journal} {Astron. Astrophys.}\ }\textbf {\bibinfo
  {volume} {641}},\ \bibinfo {pages} {A6} (\bibinfo {year}
  {2020}{\natexlab{b}})},\ \bibinfo {note} {[Erratum: Astron.Astrophys. 652, C4
  (2021)]},\ \Eprint {http://arxiv.org/abs/1807.06209} {arXiv:1807.06209
  [astro-ph.CO]} \BibitemShut {NoStop}%
\bibitem [{\citenamefont {Hazumi}\ \emph {et~al.}(2020)\citenamefont {Hazumi}
  \emph {et~al.}}]{LiteBIRD:2020khw}%
  \BibitemOpen
  \bibfield  {author} {\bibinfo {author} {\bibfnamefont {M.}~\bibnamefont
  {Hazumi}} \emph {et~al.} (\bibinfo {collaboration} {LiteBIRD}),\ }\bibfield
  {title} {\enquote {\bibinfo {title} {{LiteBIRD: JAXA's new strategic L-class
  mission for all-sky surveys of cosmic microwave background polarization}},}\
  }\href {\doibase 10.1117/12.2563050} {\bibfield  {journal} {\bibinfo
  {journal} {Proc. SPIE Int. Soc. Opt. Eng.}\ }\textbf {\bibinfo {volume}
  {11443}},\ \bibinfo {pages} {114432F} (\bibinfo {year} {2020})},\ \Eprint
  {http://arxiv.org/abs/2101.12449} {arXiv:2101.12449 [astro-ph.IM]}
  \BibitemShut {NoStop}%
\bibitem [{\citenamefont {Aghamousa}\ \emph {et~al.}(2016)\citenamefont
  {Aghamousa} \emph {et~al.}}]{DESI:2016fyo}%
  \BibitemOpen
  \bibfield  {author} {\bibinfo {author} {\bibfnamefont {Amir}\ \bibnamefont
  {Aghamousa}} \emph {et~al.} (\bibinfo {collaboration} {DESI}),\ }\bibfield
  {title} {\enquote {\bibinfo {title} {{The DESI Experiment Part I:
  Science,Targeting, and Survey Design}},}\ }\href@noop {} {\  (\bibinfo {year}
  {2016})},\ \Eprint {http://arxiv.org/abs/1611.00036} {arXiv:1611.00036
  [astro-ph.IM]} \BibitemShut {NoStop}%
\bibitem [{\citenamefont {Font-Ribera}\ \emph {et~al.}(2014)\citenamefont
  {Font-Ribera}, \citenamefont {McDonald}, \citenamefont {Mostek},
  \citenamefont {Reid}, \citenamefont {Seo},\ and\ \citenamefont
  {Slosar}}]{Font-Ribera:2013rwa}%
  \BibitemOpen
  \bibfield  {author} {\bibinfo {author} {\bibfnamefont {Andreu}\ \bibnamefont
  {Font-Ribera}}, \bibinfo {author} {\bibfnamefont {Patrick}\ \bibnamefont
  {McDonald}}, \bibinfo {author} {\bibfnamefont {Nick}\ \bibnamefont {Mostek}},
  \bibinfo {author} {\bibfnamefont {Beth~A.}\ \bibnamefont {Reid}}, \bibinfo
  {author} {\bibfnamefont {Hee-Jong}\ \bibnamefont {Seo}}, \ and\ \bibinfo
  {author} {\bibfnamefont {An}~\bibnamefont {Slosar}},\ }\bibfield  {title}
  {\enquote {\bibinfo {title} {{DESI and other dark energy experiments in the
  era of neutrino mass measurements}},}\ }\href {\doibase
  10.1088/1475-7516/2014/05/023} {\bibfield  {journal} {\bibinfo  {journal}
  {JCAP}\ }\textbf {\bibinfo {volume} {05}},\ \bibinfo {pages} {023} (\bibinfo
  {year} {2014})},\ \Eprint {http://arxiv.org/abs/1308.4164} {arXiv:1308.4164
  [astro-ph.CO]} \BibitemShut {NoStop}%
\bibitem [{\citenamefont {Allison}\ \emph {et~al.}(2015)\citenamefont
  {Allison}, \citenamefont {Caucal}, \citenamefont {Calabrese}, \citenamefont
  {Dunkley},\ and\ \citenamefont {Louis}}]{Allison:2015qca}%
  \BibitemOpen
  \bibfield  {author} {\bibinfo {author} {\bibfnamefont {R.}~\bibnamefont
  {Allison}}, \bibinfo {author} {\bibfnamefont {P.}~\bibnamefont {Caucal}},
  \bibinfo {author} {\bibfnamefont {E.}~\bibnamefont {Calabrese}}, \bibinfo
  {author} {\bibfnamefont {J.}~\bibnamefont {Dunkley}}, \ and\ \bibinfo
  {author} {\bibfnamefont {T.}~\bibnamefont {Louis}},\ }\bibfield  {title}
  {\enquote {\bibinfo {title} {{Towards a cosmological neutrino mass
  detection}},}\ }\href {\doibase 10.1103/PhysRevD.92.123535} {\bibfield
  {journal} {\bibinfo  {journal} {Phys. Rev. D}\ }\textbf {\bibinfo {volume}
  {92}},\ \bibinfo {pages} {123535} (\bibinfo {year} {2015})},\ \Eprint
  {http://arxiv.org/abs/1509.07471} {arXiv:1509.07471 [astro-ph.CO]}
  \BibitemShut {NoStop}%
\bibitem [{\citenamefont {Millea}\ and\ \citenamefont
  {Seljak}(2021)}]{Millea:2021had}%
  \BibitemOpen
  \bibfield  {author} {\bibinfo {author} {\bibfnamefont {Marius}\ \bibnamefont
  {Millea}}\ and\ \bibinfo {author} {\bibfnamefont {Uros}\ \bibnamefont
  {Seljak}},\ }\bibfield  {title} {\enquote {\bibinfo {title} {{MUSE: Marginal
  Unbiased Score Expansion and Application to CMB Lensing}},}\ }\href@noop {}
  {\  (\bibinfo {year} {2021})},\ \Eprint {http://arxiv.org/abs/2112.09354}
  {arXiv:2112.09354 [astro-ph.CO]} \BibitemShut {NoStop}%
\bibitem [{\citenamefont {van Engelen}\ \emph {et~al.}(2014)\citenamefont {van
  Engelen}, \citenamefont {Bhattacharya}, \citenamefont {Sehgal}, \citenamefont
  {Holder}, \citenamefont {Zahn},\ and\ \citenamefont
  {Nagai}}]{vanEngelen:2013rla}%
  \BibitemOpen
  \bibfield  {author} {\bibinfo {author} {\bibfnamefont {A.}~\bibnamefont {van
  Engelen}}, \bibinfo {author} {\bibfnamefont {S.}~\bibnamefont
  {Bhattacharya}}, \bibinfo {author} {\bibfnamefont {N.}~\bibnamefont
  {Sehgal}}, \bibinfo {author} {\bibfnamefont {G.~P.}\ \bibnamefont {Holder}},
  \bibinfo {author} {\bibfnamefont {O.}~\bibnamefont {Zahn}}, \ and\ \bibinfo
  {author} {\bibfnamefont {D.}~\bibnamefont {Nagai}},\ }\bibfield  {title}
  {\enquote {\bibinfo {title} {{CMB Lensing Power Spectrum Biases from Galaxies
  and Clusters using High-angular Resolution Temperature Maps}},}\ }\href
  {\doibase 10.1088/0004-637X/786/1/13} {\bibfield  {journal} {\bibinfo
  {journal} {Astrophys. J.}\ }\textbf {\bibinfo {volume} {786}},\ \bibinfo
  {pages} {13} (\bibinfo {year} {2014})},\ \Eprint
  {http://arxiv.org/abs/1310.7023} {arXiv:1310.7023 [astro-ph.CO]} \BibitemShut
  {NoStop}%
\bibitem [{\citenamefont {Osborne}\ \emph {et~al.}(2014)\citenamefont
  {Osborne}, \citenamefont {Hanson},\ and\ \citenamefont
  {Dor\'e}}]{Osborne:2013nna}%
  \BibitemOpen
  \bibfield  {author} {\bibinfo {author} {\bibfnamefont {Stephen~J.}\
  \bibnamefont {Osborne}}, \bibinfo {author} {\bibfnamefont {Duncan}\
  \bibnamefont {Hanson}}, \ and\ \bibinfo {author} {\bibfnamefont {Olivier}\
  \bibnamefont {Dor\'e}},\ }\bibfield  {title} {\enquote {\bibinfo {title}
  {{Extragalactic Foreground Contamination in Temperature-based CMB Lens
  Reconstruction}},}\ }\href {\doibase 10.1088/1475-7516/2014/03/024}
  {\bibfield  {journal} {\bibinfo  {journal} {JCAP}\ }\textbf {\bibinfo
  {volume} {03}},\ \bibinfo {pages} {024} (\bibinfo {year} {2014})},\ \Eprint
  {http://arxiv.org/abs/1310.7547} {arXiv:1310.7547 [astro-ph.CO]} \BibitemShut
  {NoStop}%
\bibitem [{\citenamefont {Ferraro}\ and\ \citenamefont
  {Hill}(2018)}]{Ferraro:2017fac}%
  \BibitemOpen
  \bibfield  {author} {\bibinfo {author} {\bibfnamefont {Simone}\ \bibnamefont
  {Ferraro}}\ and\ \bibinfo {author} {\bibfnamefont {J.~Colin}\ \bibnamefont
  {Hill}},\ }\bibfield  {title} {\enquote {\bibinfo {title} {{Bias to CMB
  Lensing Reconstruction from Temperature Anisotropies due to Large-Scale
  Galaxy Motions}},}\ }\href {\doibase 10.1103/PhysRevD.97.023512} {\bibfield
  {journal} {\bibinfo  {journal} {Phys. Rev. D}\ }\textbf {\bibinfo {volume}
  {97}},\ \bibinfo {pages} {023512} (\bibinfo {year} {2018})},\ \Eprint
  {http://arxiv.org/abs/1705.06751} {arXiv:1705.06751 [astro-ph.CO]}
  \BibitemShut {NoStop}%
\bibitem [{\citenamefont {Madhavacheril}\ and\ \citenamefont
  {Hill}(2018)}]{Madhavacheril:2018bxi}%
  \BibitemOpen
  \bibfield  {author} {\bibinfo {author} {\bibfnamefont {Mathew~S.}\
  \bibnamefont {Madhavacheril}}\ and\ \bibinfo {author} {\bibfnamefont
  {J.~Colin}\ \bibnamefont {Hill}},\ }\bibfield  {title} {\enquote {\bibinfo
  {title} {{Mitigating Foreground Biases in CMB Lensing Reconstruction Using
  Cleaned Gradients}},}\ }\href {\doibase 10.1103/PhysRevD.98.023534}
  {\bibfield  {journal} {\bibinfo  {journal} {Phys. Rev. D}\ }\textbf {\bibinfo
  {volume} {98}},\ \bibinfo {pages} {023534} (\bibinfo {year} {2018})},\
  \Eprint {http://arxiv.org/abs/1802.08230} {arXiv:1802.08230 [astro-ph.CO]}
  \BibitemShut {NoStop}%
\bibitem [{\citenamefont {Schaan}\ and\ \citenamefont
  {Ferraro}(2019)}]{Schaan:2018tup}%
  \BibitemOpen
  \bibfield  {author} {\bibinfo {author} {\bibfnamefont {Emmanuel}\
  \bibnamefont {Schaan}}\ and\ \bibinfo {author} {\bibfnamefont {Simone}\
  \bibnamefont {Ferraro}},\ }\bibfield  {title} {\enquote {\bibinfo {title}
  {{Foreground-Immune Cosmic Microwave Background Lensing with Shear-Only
  Reconstruction}},}\ }\href {\doibase 10.1103/PhysRevLett.122.181301}
  {\bibfield  {journal} {\bibinfo  {journal} {Phys. Rev. Lett.}\ }\textbf
  {\bibinfo {volume} {122}},\ \bibinfo {pages} {181301} (\bibinfo {year}
  {2019})},\ \Eprint {http://arxiv.org/abs/1804.06403} {arXiv:1804.06403
  [astro-ph.CO]} \BibitemShut {NoStop}%
\bibitem [{\citenamefont {Mishra}\ and\ \citenamefont
  {Schaan}(2019)}]{Mishra:2019qyd}%
  \BibitemOpen
  \bibfield  {author} {\bibinfo {author} {\bibfnamefont {Nishant}\ \bibnamefont
  {Mishra}}\ and\ \bibinfo {author} {\bibfnamefont {Emmanuel}\ \bibnamefont
  {Schaan}},\ }\bibfield  {title} {\enquote {\bibinfo {title} {{Bias to CMB
  lensing from lensed foregrounds}},}\ }\href {\doibase
  10.1103/PhysRevD.100.123504} {\bibfield  {journal} {\bibinfo  {journal}
  {Phys. Rev. D}\ }\textbf {\bibinfo {volume} {100}},\ \bibinfo {pages}
  {123504} (\bibinfo {year} {2019})},\ \Eprint
  {http://arxiv.org/abs/1908.08057} {arXiv:1908.08057 [astro-ph.CO]}
  \BibitemShut {NoStop}%
\bibitem [{\citenamefont {Sailer}\ \emph {et~al.}(2020)\citenamefont {Sailer},
  \citenamefont {Schaan},\ and\ \citenamefont {Ferraro}}]{Sailer:2020lal}%
  \BibitemOpen
  \bibfield  {author} {\bibinfo {author} {\bibfnamefont {Noah}\ \bibnamefont
  {Sailer}}, \bibinfo {author} {\bibfnamefont {Emmanuel}\ \bibnamefont
  {Schaan}}, \ and\ \bibinfo {author} {\bibfnamefont {Simone}\ \bibnamefont
  {Ferraro}},\ }\bibfield  {title} {\enquote {\bibinfo {title} {{Lower bias,
  lower noise CMB lensing with foreground-hardened estimators}},}\ }\href
  {\doibase 10.1103/PhysRevD.102.063517} {\bibfield  {journal} {\bibinfo
  {journal} {Phys. Rev. D}\ }\textbf {\bibinfo {volume} {102}},\ \bibinfo
  {pages} {063517} (\bibinfo {year} {2020})},\ \Eprint
  {http://arxiv.org/abs/2007.04325} {arXiv:2007.04325 [astro-ph.CO]}
  \BibitemShut {NoStop}%
\bibitem [{\citenamefont {Darwish}\ \emph {et~al.}(2021)\citenamefont
  {Darwish}, \citenamefont {Sherwin}, \citenamefont {Sailer}, \citenamefont
  {Schaan},\ and\ \citenamefont {Ferraro}}]{Darwish:2021ycf}%
  \BibitemOpen
  \bibfield  {author} {\bibinfo {author} {\bibfnamefont {Omar}\ \bibnamefont
  {Darwish}}, \bibinfo {author} {\bibfnamefont {Blake~D.}\ \bibnamefont
  {Sherwin}}, \bibinfo {author} {\bibfnamefont {Noah}\ \bibnamefont {Sailer}},
  \bibinfo {author} {\bibfnamefont {Emmanuel}\ \bibnamefont {Schaan}}, \ and\
  \bibinfo {author} {\bibfnamefont {Simone}\ \bibnamefont {Ferraro}},\
  }\bibfield  {title} {\enquote {\bibinfo {title} {{Optimizing foreground
  mitigation for CMB lensing with combined multifrequency and geometric
  methods}},}\ }\href@noop {} {\  (\bibinfo {year} {2021})},\ \Eprint
  {http://arxiv.org/abs/2111.00462} {arXiv:2111.00462 [astro-ph.CO]}
  \BibitemShut {NoStop}%
\bibitem [{\citenamefont {Lembo}\ \emph {et~al.}(2021)\citenamefont {Lembo},
  \citenamefont {Fabbian}, \citenamefont {Carron},\ and\ \citenamefont
  {Lewis}}]{Lembo:2021kxc}%
  \BibitemOpen
  \bibfield  {author} {\bibinfo {author} {\bibfnamefont {Margherita}\
  \bibnamefont {Lembo}}, \bibinfo {author} {\bibfnamefont {Giulio}\
  \bibnamefont {Fabbian}}, \bibinfo {author} {\bibfnamefont {Julien}\
  \bibnamefont {Carron}}, \ and\ \bibinfo {author} {\bibfnamefont {Antony}\
  \bibnamefont {Lewis}},\ }\bibfield  {title} {\enquote {\bibinfo {title} {{CMB
  lensing reconstruction biases from masking extragalactic sources}},}\
  }\href@noop {} {\  (\bibinfo {year} {2021})},\ \Eprint
  {http://arxiv.org/abs/2109.13911} {arXiv:2109.13911 [astro-ph.CO]}
  \BibitemShut {NoStop}%
\bibitem [{\citenamefont {Hirata}\ \emph {et~al.}(2008)\citenamefont {Hirata},
  \citenamefont {Ho}, \citenamefont {Padmanabhan}, \citenamefont {Seljak},\
  and\ \citenamefont {Bahcall}}]{Hirata:2008cb}%
  \BibitemOpen
  \bibfield  {author} {\bibinfo {author} {\bibfnamefont {Christopher~M.}\
  \bibnamefont {Hirata}}, \bibinfo {author} {\bibfnamefont {Shirley}\
  \bibnamefont {Ho}}, \bibinfo {author} {\bibfnamefont {Nikhil}\ \bibnamefont
  {Padmanabhan}}, \bibinfo {author} {\bibfnamefont {Uros}\ \bibnamefont
  {Seljak}}, \ and\ \bibinfo {author} {\bibfnamefont {Neta~A.}\ \bibnamefont
  {Bahcall}},\ }\bibfield  {title} {\enquote {\bibinfo {title} {{Correlation of
  CMB with large-scale structure: II. Weak lensing}},}\ }\href {\doibase
  10.1103/PhysRevD.78.043520} {\bibfield  {journal} {\bibinfo  {journal} {Phys.
  Rev. D}\ }\textbf {\bibinfo {volume} {78}},\ \bibinfo {pages} {043520}
  (\bibinfo {year} {2008})},\ \Eprint {http://arxiv.org/abs/0801.0644}
  {arXiv:0801.0644 [astro-ph]} \BibitemShut {NoStop}%
\bibitem [{\citenamefont {Schaan}\ \emph {et~al.}(2017)\citenamefont {Schaan},
  \citenamefont {Krause}, \citenamefont {Eifler}, \citenamefont {Dor\'e},
  \citenamefont {Miyatake}, \citenamefont {Rhodes},\ and\ \citenamefont
  {Spergel}}]{Schaan:2016ois}%
  \BibitemOpen
  \bibfield  {author} {\bibinfo {author} {\bibfnamefont {Emmanuel}\
  \bibnamefont {Schaan}}, \bibinfo {author} {\bibfnamefont {Elisabeth}\
  \bibnamefont {Krause}}, \bibinfo {author} {\bibfnamefont {Tim}\ \bibnamefont
  {Eifler}}, \bibinfo {author} {\bibfnamefont {Olivier}\ \bibnamefont
  {Dor\'e}}, \bibinfo {author} {\bibfnamefont {Hironao}\ \bibnamefont
  {Miyatake}}, \bibinfo {author} {\bibfnamefont {Jason}\ \bibnamefont
  {Rhodes}}, \ and\ \bibinfo {author} {\bibfnamefont {David~N.}\ \bibnamefont
  {Spergel}},\ }\bibfield  {title} {\enquote {\bibinfo {title} {{Looking
  through the same lens: Shear calibration for LSST, Euclid, and WFIRST with
  stage 4 CMB lensing}},}\ }\href {\doibase 10.1103/PhysRevD.95.123512}
  {\bibfield  {journal} {\bibinfo  {journal} {Phys. Rev. D}\ }\textbf {\bibinfo
  {volume} {95}},\ \bibinfo {pages} {123512} (\bibinfo {year} {2017})},\
  \Eprint {http://arxiv.org/abs/1607.01761} {arXiv:1607.01761 [astro-ph.CO]}
  \BibitemShut {NoStop}%
\bibitem [{\citenamefont {Schmittfull}\ and\ \citenamefont
  {Seljak}(2018)}]{Schmittfull:2017ffw}%
  \BibitemOpen
  \bibfield  {author} {\bibinfo {author} {\bibfnamefont {Marcel}\ \bibnamefont
  {Schmittfull}}\ and\ \bibinfo {author} {\bibfnamefont {Uros}\ \bibnamefont
  {Seljak}},\ }\bibfield  {title} {\enquote {\bibinfo {title} {{Parameter
  constraints from cross-correlation of CMB lensing with galaxy clustering}},}\
  }\href {\doibase 10.1103/PhysRevD.97.123540} {\bibfield  {journal} {\bibinfo
  {journal} {Phys. Rev. D}\ }\textbf {\bibinfo {volume} {97}},\ \bibinfo
  {pages} {123540} (\bibinfo {year} {2018})},\ \Eprint
  {http://arxiv.org/abs/1710.09465} {arXiv:1710.09465 [astro-ph.CO]}
  \BibitemShut {NoStop}%
\bibitem [{\citenamefont {Abbott}\ \emph {et~al.}(2019)\citenamefont {Abbott}
  \emph {et~al.}}]{DES:2018fvb}%
  \BibitemOpen
  \bibfield  {author} {\bibinfo {author} {\bibfnamefont {T.~M.~C.}\
  \bibnamefont {Abbott}} \emph {et~al.} (\bibinfo {collaboration} {DES, SPT}),\
  }\bibfield  {title} {\enquote {\bibinfo {title} {{Dark Energy Survey Year 1
  Results: Joint Analysis of Galaxy Clustering, Galaxy Lensing, and CMB Lensing
  Two-point Functions}},}\ }\href {\doibase 10.1103/PhysRevD.100.023541}
  {\bibfield  {journal} {\bibinfo  {journal} {Phys. Rev. D}\ }\textbf {\bibinfo
  {volume} {100}},\ \bibinfo {pages} {023541} (\bibinfo {year} {2019})},\
  \Eprint {http://arxiv.org/abs/1810.02322} {arXiv:1810.02322 [astro-ph.CO]}
  \BibitemShut {NoStop}%
\bibitem [{\citenamefont {Ili\'c}\ \emph {et~al.}(2021)\citenamefont {Ili\'c}
  \emph {et~al.}}]{Euclid:2021qvm}%
  \BibitemOpen
  \bibfield  {author} {\bibinfo {author} {\bibfnamefont {S.}~\bibnamefont
  {Ili\'c}} \emph {et~al.} (\bibinfo {collaboration} {Euclid}),\ }\bibfield
  {title} {\enquote {\bibinfo {title} {{$Euclid$ preparation: XV. Forecasting
  cosmological constraints for the $Euclid$ and CMB joint analysis}},}\ }\href
  {\doibase 10.1051/0004-6361/202141556} {\  (\bibinfo {year} {2021}),\
  10.1051/0004-6361/202141556},\ \Eprint {http://arxiv.org/abs/2106.08346}
  {arXiv:2106.08346 [astro-ph.CO]} \BibitemShut {NoStop}%
\end{thebibliography}%

\end{document}